\documentclass[aps,pra,twocolumn,floatfix,10pt,notitlepage,showpacs]{revtex4-1} 

\usepackage{amsmath,amsfonts,amssymb,graphicx,bbm,times}
\usepackage{dsfont}

\usepackage{listings}

\usepackage[usenames,dvipsnames]{xcolor}   
\usepackage{ulem}                          

\usepackage{hyperref}

\usepackage[utf8]{inputenc}

\DeclareMathOperator{\tr}{tr}

\hypersetup{colorlinks=true,
						linkcolor=Blue,        
						citecolor=Blue,        
						filecolor=blue,      			 
						urlcolor=black                  
}

\preprint{\bf PREPRINT}

\begin{document}

\bibliographystyle{apsrev}
\newtheorem{example}{Example}

\newtheorem{remark}{Remark}
\newtheorem{problem}{Problem}
\newtheorem{theorem}{Theorem}
\newtheorem{corollary}{Corollary}
\newtheorem{definition}{Definition}
\newtheorem{proposition}{Proposition}
\newtheorem{lemma}{Lemma}
\newtheorem{conjecture}{Conjecture}
\newcommand{\proofend}{\hfill\fbox\\\medskip }
\newcommand{\proof}[1]{{\bf{Proof.}} #1 $\proofend$}
\newcommand{\nn}{{\mathbbm{N}}}
\newcommand{\rr}{{\mathbbm{R}}}
\newcommand{\cc}{{\mathbbm{C}}}
\newcommand{\zz}{{\mathbbm{Z}}}
\newcommand{\mbp}{\ensuremath{\spadesuit}}
\newcommand{\je}{\ensuremath{\heartsuit}}
\newcommand{\jd}{\ensuremath{\clubsuit}}
\newcommand{\id}{{\mathbbm{1}}}
\renewcommand{\vec}[1]{\boldsymbol{#1}}
\newcommand{\me}{\mathrm{e}}
\newcommand{\mi}{\mathrm{i}}
\newcommand{\md}{\mathrm{d}}
\newcommand{\sg}{\text{sgn}}
\newcommand{\mc}[1]{\textcolor{blue}{[#1]}}

\def\>{\rangle}
\def\<{\langle}
\def\({\left(}
\def\){\right)}

\newcommand{\ket}[1]{\left|#1\right>}
\newcommand{\bra}[1]{\left<#1\right|}
\newcommand{\braket}[2]{\<#1|#2\>}
\newcommand{\ketbra}[2]{\left|#1\right>\!\left<#2\right|}
\newcommand{\proj}[1]{|#1\>\!\<#1|}
\newcommand{\avg}[1]{\< #1 \>}

\renewcommand{\tensor}{\otimes}

\newcommand{\einfuegen}[1]{\textcolor{PineGreen}{#1}}
\newcommand{\streichen}[1]{\textcolor{red}{\sout{#1}}}
\newcommand{\todo}[1]{\textcolor{blue}{(ToDo: #1)}}
\newcommand{\transpose}[1]{{#1}^t}

\newcommand{\om}[1]{\textcolor{black}{#1}}


\title{Multiparticle entanglement criteria for nonsymmetric collective variances}

\author{O.\ Marty$^1$, M.\ Cramer$^{1,2}$, G.\ Vitagliano$^3$, G.\ T\'oth$^{4,5,6}$, and M.B.\ Plenio$^1$}
\affiliation{
$^1$Institut f\"ur Theoretische Physik, Universit\"at Ulm, D-89081 Ulm, Germany\\
$^2$Institut für Theoretische Physik, Leibniz Universität Hannover, D-30167 Hannover, Germany \\
$^3$Institute for Quantum Optics and Quantum Information (IQOQI), Austrian Academy of Sciences, A-1090 Vienna, Austria \\
$^4$Department of Theoretical Physics, University of the Basque Country UPV/EHU, P.O.\ Box 644, E-48080 Bilbao, Spain \\
$^5$IKERBASQUE, Basque Foundation for Science, E-48013 Bilbao, Spain \\
$^6$Wigner Research Centre for Physics, Hungarian Academy of Sciences, P.O.\ Box 49, H-1525 Budapest, Hungary}

\begin{abstract}
We introduce a general scheme to detect various multiparticle entanglement structures from global non-permutationally invariant observables. In particular, we derive bounds on the variance of non-permutationally invariant and collective operators for the verification of $k$-party entanglement. 
For a family of observables related to the spin structure factor, we give quantitative bounds on entanglement that are independent of the total number of particles. We introduce highly non-symmetric states with genuine multipartite entanglement that is verifiable with the presented technique and discuss how they can be prepared with trapped ions exploiting the high degree of control in these systems.
As a special case, our framework provides an alternative approach to obtain a tight relaxation of the entanglement criterion by S\o rensen and M\o lmer [Phys.\ Rev.\ Lett.\ 86, 4431 (2001)] that is free from technical assumptions and allows to calculate the bounds with an improved scaling in the detectable depth.
\end{abstract}


\date{\today}

\maketitle

\section{Introduction} 
One of the most fascinating challenges in quantum information science is to explore the prospects of quantum effects to go beyond the capabilities of classical physics. An example is the concept of spin-squeezing that describes a collective property of an aggregation of spins \cite{Wineland,Kitagawa}. Originally, it has been introduced to achieve performance gain using quantum metrology.  The basic concept can be quantified in a multitude of spin-squeezing parameters \cite{Wineland,Kitagawa,Ma}. \om{The role of spin-squeezing in the context of quantum improved measurements can be illustrated graphically, providing an intuitive picture of the spin-squeezing parameters \cite{Kitagawa,Ma}.}

On the other hand, a different central application of spin squeezing parameters is the detection of many-body quantum correlations: strongly (anti-)correlated spins in squeezed states exceeding the standard quantum limit are required to be entangled as has been observed in \cite{Korbicz, Toth2009, sorensen}.
Quantitatively the degree of spin-squeezing is a measure of multiparty entanglement, i.e.\ in a strongly squeezed state the entanglement necessarily spreads among a large number of spins.
 Entanglement criteria based on spin-squeezing parameters benefit from the fact that these parameters usually depend on simple and global observables only, in particular, typically on low-order moments of collective spin operators. 
 
 This has two major implications. Experimentally, the approach provides an accessible and robust way for entanglement detection that is free of any assumptions on the system and may therefore be suitable for many different platforms. On the theoretical side, a criterion may be obtained from two main ingredients: (i) local uncertainty relations due to the few-body correlations involved and (ii) exploiting the permutation invariance of the observables. As a consequence, this reduces the complexity of the task of determining entanglement criteria drastically and hence, for example, a complete set of inequalities useful for the detection of non-separability for the first and second moments of the magnetization may be given explicitly \cite{Vitagliano}. Yet, these simplifications also set the limitations to the spin-squeezing criteria. Extending them is therefore desirable, \om{in particular, to platforms with non-permutation invariant observables \cite{blatt, bloch, devoret}, to open them up to entanglement schemes that are established in the permutation invariant setting.}
 
 In this work, we focus on criteria which do not rely on permutation invariant observables. To this end, it is important to note that the methodology of entanglement detection via spin-squeezing, in particular the application of local uncertainty relations, is also applicable to observable quantities other than collective spin operators \cite{Vitagliano3}. Here, we apply Lagrange-duality to a specific constrained optimization problem to introduce a general scheme which allows for the detection of many-body entanglement via global observables. To give a concrete example, we will focus on Fourier-transformed spin operators. Such observables arise in scattering experiments and they are intrinsically non-permutation invariant. A prominent example is the static structure factor, which is accessible, e.g., by neutron scattering from magnetic materials. The structure factor has been demonstrated useful for entanglement quantification for example in the vicinity of phased Dicke states \cite{cramer,marty,krammer} where usual spin-squeezing criteria are unable to confirm entanglement. \om{The method presented hereinafter can be used to detect $k$-party and other forms of multipartite entanglement such as $k$-wide entanglement by means of the structure factor  \cite{Woelk}.} \om{To this end, the route of Lagrange duality turns out to be a fruitful way of approaching the problem.}
 	
 \om{Strikingly, this approach provides an alternative way to derive the multiparty entanglement criteria of Ref.\ \cite{sorensen} if it is combined with a numerically efficient method for lower bounds to ground state energies introduced by Baumgratz and Plenio in \cite{baumgratz}. With this, it is possible to close gaps in the proof of the entanglement bounds \cite{sorensen} and, moreover, under a mild (and numerically testable) assumption that has also been exploited in Ref.\ \cite{sorensen} we can calculate the bounds for any entanglement depth $k$ without the need of increasing the Hilbert space dimension of the underlying problem. This impacts existing experimental and theoretical work that builds on Ref.\ \cite{sorensen}, see e.g.\ \cite{Vitagliano2, luecke, gross,hosten,riedel,cox,engels,dellantonio}, and paves the way for the detection of larger, potentially macroscopic, numbers of entangled particles.} 

For the more general non-permutation symmetric observables, in order to calculate the criteria explicitly, we consider an algorithm for global, non-convex eigenvalue optimization which could be combined with matrix-product state methods. The general case is of practical interest as, e.g., they may be accessible in scattering experiments with neutrons on crystalline magnetic compounds \cite{Jensen} or with X-Ray light on cold atoms \cite{hart,mazurenko}. Finally, we construct states, that can be proven to be genuine multipartite entangled by our scheme, by demonstrating how they can experimentally be generated with trapped ions using the high degree of control over the interaction provided by these systems. These findings support the versatility and practical importance of our framework for the field of controlled quantum systems.

\section{Preliminaries}

In order to obtain a detection scheme for $k$-party entanglement, S\o rensen and M\o lmer determined the minimal variance of the collective spin operator of a many-body state as a function of its magnetization in one of the orthogonal directions. 
To generalize these results we start by introducing the variance of a (not necessarily Hermitian) operator $\hat O$ in a state $\hat \varrho$ as
\begin{equation}
\label{gen_var}
\Delta_{\hat\varrho}^2[\hat O ] := \langle \hat O^\dag \hat O \rangle_{\hat\varrho} - \langle \hat O^\dagger \rangle_{\hat\varrho}\langle \hat O \rangle_{\hat\varrho}.
\end{equation}
Eq.\ \eqref{gen_var} reduces to the usual definition of the variance if $\hat O$ is Hermitian.

The goal is now to find a function $F_{\mathcal{C}}$ such that for states $\hat \varrho$ belonging to a certain class $\mathcal{C}$ of states (e.g., $k$-producible states) one has
\begin{equation}
\label{main inequality}
\Delta_{\hat\varrho}^2 [\hat O] \ge F_{\mathcal{C}}(\langle \hat M\rangle_{\hat\varrho}),
\end{equation}
i.e.\ a lower bound to the variance in terms of an additional observable $\hat M$ playing the role of the magnetization in \cite{sorensen}.
Let us assume that we have access to $\Delta_{\varrho}^2 [\hat O]$ and $\langle \hat M\rangle_{\hat\varrho}$ in an experiment. If the measurements happen to violate the above inequality then it is guaranteed that the state in the laboratory is {\it not} in that class. E.g., if $\mathcal{C}$ corresponds to the set of $k$-producible states then such a violation shows that the state is $(k+1)$-party entangled. We set out to determine $F_{\mathcal{C}}$ for different classes $\mathcal{C}$. We start with considering $N$ spin-$S$
particles and later, for concrete examples, focus on spin chains
and $S = 1/2$. We will define classes of states as follows.

Any state $\hat \varrho$ on $N$ spins may be written as
\begin{equation}
\label{factorized state}
\hat\varrho=\sum_np_n\,\hat\varrho_1^{(n)}\!\!\otimes\cdots\otimes\!\hat\varrho_{P_n}^{(n)},\;\;\;p_n\ge 0,\;\;\;\sum_np_n=1,
\end{equation}
where each $\hat\varrho_p^{(n)}$ corresponds to the state on a subset $\mathcal{Z}^{(n)}_p$ of the $N$ spins and $P_n$ denotes the number of factors in the $n$'th summand. One may now define classes of states by restricting the subsets $\mathcal{Z}^{(n)}_p$: E.g., if one demands that all $|\mathcal{Z}^{(n)}_p|=1$ then this defines the fully separable states. If one demands $P_n=k$ then all such states are $k$-separable. If one restricts the $\mathcal{Z}^{(n)}_p$ to contain at most $k$ spins then this defines the set of $k$-producible states.

\section{Main Observation}
Consider now a certain class of states on $N$ spins, i.e., all density matrices $\hat\varrho$  as in Eq.~\eqref{factorized state} with $\mathcal{Z}^{(n)}_p\in\mathcal{C}$, where $\mathcal{C}$ defines the class under consideration. Furthermore, let $\hat O=\sum_{i=1}^N \hat O_i$ and $\hat M=\sum_{i=1}^N \hat M_i$ with $\hat O_i$ and $\hat M_i$ acting only on the $i$'th spin but potentially different operators at each $i$, i.e., we do not demand $\hat O$ nor $\hat M$ to be permutation invariant. 
Our main observation is that for {\it any} such operators one obtains (via Lagrange duality and the variational characterization of the variance \cite{damm} generalized to non-Hermitian operator, see Appendix \ref{app:main observation} for details)
a lower bound as in Eq.~\eqref{main inequality} with
\begin{equation}
\label{F_C}
\begin{split}
F_{\mathcal{C}}(m)=\sup_{\lambda\in\rr}\left(\lambda m+N\min_{\mathcal{Z}\in\mathcal{C}}\frac{G_{\mathcal{Z}}(\lambda)}{|\mathcal{Z}|}\right),
\end{split}
\end{equation}
where
\begin{equation}
\label{G_Z}
G_{\mathcal{Z}}(\lambda) = \inf_{s\in\cc}\lambda_{\text{min}} \left[ (\hat O_{\mathcal{Z}}-s\id)^\dagger (\hat O_{\mathcal{Z}}-s\id)-\lambda \hat M_{\mathcal{Z}}\right].
\end{equation}
Here, $\lambda_{\text{min}}[\cdot]$ denotes the smallest eigenvalue of the Hermitian matrix in brackets and $\hat O_{\mathcal{Z}}=\sum_{i\in\mathcal{Z}}\hat O_i$ and similarly for $\hat M$. We note that $G_{\mathcal{Z}}$ is concave and $F_{\mathcal{C}}$ is convex. Furthermore, the above holds for any collection of spins, such that, e.g. $D$-dimensional lattices are included. If the variance of $\hat O$ and the mean value of $\hat M$ are experimentally accessible and violate the inequality in Eq.~\eqref{main inequality} then, without making any further assumptions, one can conclude that the state in the laboratory is not in the class $\mathcal{C}$. How strong the bound can be violated depends on the observables. \om{So far, we have introduced a framework for the detection of various multiparticle entanglement structures by global measurements without making any assumption on the underlying system. As the violation of these criteria necessitates a sufficiently small variance Eq.\ \eqref{gen_var} it may be seen as a generalization of the spin-squeezing phenomenon to, both, arbitrary observables and more general forms of entanglement.}

Notably, we only require that the operators are the sum of single site operators, so that, e.g., $\hat O = \sum_{i=1}^N f_i \hat\sigma_z^{i}$ with $f_i \in \cc$ fits into our framework. Additionally, a bound to a sum of variances in terms of the expectation values of multiple observables can directly be incorporated into equation Eq.\ \eqref{F_C}, if these operators have the above local form \cite{vitagliano3}.

Now, how hard is it to actually compute $F_{\mathcal{C}}$? First of all, to determine $G_{\mathcal{Z}}$, {(a)} one needs to be able to find the smallest eigenvalue of a potentially very large matrix: a priori, the dimensions of the involved matrices are exponentially large in $|\mathcal{Z}|$. If, e.g., the goal is to detect $k$-party entanglement then the dimensions of the involved matrices are exponentially large in $k$.

Secondly, in order to obtain $F_{\mathcal{C}}$, {(b)} one needs to determine $G_{\mathcal{Z}}$ for {\it all} $\mathcal{Z}\in\mathcal{C}$. Considering the example of $k$-party entanglement again, one needs to compute $G_{\mathcal{Z}}$ for {\it all} subsets $\mathcal{Z}$ containing at most $k-1$ spins as we allow for non-permutation-invariant operators. If the involved operators are permutation invariant this complexity is dramatically reduced: It is then sufficient to determine  $G_{\mathcal{Z}}$ for $\mathcal{Z}=\{1\},\{1,2\},\dots,\{1,\dots,k-1\}$. 

Finally, {(c)} the function to be minimized over $s$ in Eq.~\eqref{G_Z} may exhibit local minima and thus calls for global non-convex optimization.

 We remark that {(a)} may be addressed using efficient methods such as the density-matrix renormalization group (DMRG) exploiting the local form of the observables. Importantly, to guarantee a lower bound one has to carefully monitor convergence. On the other hand, it is also possible to utilize a scheme based on a semi definite program (SDP) that provides a lower bound to $\lambda_{\min}(s)$ \cite{baumgratz}. Notably, for the permutationally invariant case where $\hat O$ and $\hat M$ are given by collective spin-$1/2$ operators in two orthogonal directions, we find a relaxation of the SDP that calculates a lower bound to the smallest eigenvalue \om{where the size of the configuration $|\mathcal{Z}|$ enters the optimization soley as a parameter} and present an application below. The technical details of this method are shifted to Appendix \ref{app:lower bound}.
 
The third step, {(c)}, can be tackled utilizing an algorithm introduced in Ref.~\cite{mengi} that is based on quadratic support functions, which are determined by (i) the value of the eigenvalue function, (ii) its derivative for specific values of $s$, and (iii) an estimate of the curvature that is given as an input to the algorithm. To reliably obtain a global optimum, the estimate of the curvature is required to be a lower bound on the second derivative of the eigenvalue function in the entire parameter range. Here, we use the algorithm heuristically, decreasing the estimated curvature until we do not observe any change in the result, see Appendix \ref{Algorithm} for details. Note, that in order to provide results for large numbers of spins in a chain, one may combine this algorithm with matrix-product states (MPSs) and operators (MPOs) by reformulating steps (i) and (ii) in terms of MPSs and MPOs. As mentioned above, step (i) is a simple ground state search as can be carried out using DMRG. Once the eigenvalue function has been evaluated using DMRG, the derivative may be obtained from the calculated optimal state, see Appendix \ref{Algorithm}. \om{As a general algorithm for eigenvalue optimization of matrix-valued functions, it might be a useful tool also for other application in quantum science.}

On the other hand, when we use in step {(a)} the scheme of Ref.\ \cite{baumgratz} to obtain a lower bound to the lowest eigenvalue, the minimization over $s$ is carried out by computing the function in the entire parameter range, supported by the standard optimization toolbox of MATLAB. \om{For the permutation invariant collective spin observables we emphasize again that the size of the configuration $|\mathcal{Z}|$ enters the optimization as a parameter only.}

\section{Witnessing Multiparty Entanglement}

A state is $k$-producible if it can be decomposed as in Eq.\ \eqref{factorized state} with each $\hat\varrho_p^{(n)}$ corresponding to a state of at most $k$ spins. Denoting by $[N]=\{1,\dots,N\}$ the set of all spins, we hence have that $k$-producible states fulfil Eq.~\eqref{main inequality} with $F_{\mathcal{C}}$ as in Eq.~\eqref{F_C} and
\begin{equation}
\label{k producible C}
\mathcal{C}=\left\{\mathcal{Z}\subset [N]\,\big|\,|\mathcal{Z}|\le k\right\}.
\end{equation}
Operationally, a pure $k$-producible state can be prepared from a fully separable state via an interaction that acts on non-overlapping sets of spins separately and the number of spins in each set is upper bounded by $k$. Mixed $k$-producible states are just mixtures of the states of the above type. States which are not $k$-producible are $(k+1)$-party entangled. \om{Experimentally, in different setups the presence of $k$-party entanglement has been verified, see e.g.\ \cite{luecke, monz, barreiro, haas, mcconnell, gross, hosten, mazurenko, riedel, hart, cox, engels, dellantonio}}. $k$-party entanglement provides a multiparticle entanglement hierarchy that can be verified for a large system even if only a subset of spins can be accessed: the number of parties that may be confirmed to be entangled on the subset also gives a lower bound to the entanglement of full system \cite{guehne}. In contrast, $k$-partite entanglement cannot solely be specified by conditions on each subset $\mathcal{Z}_p^{(n)}$ separately, but needs knowledge of the full system that one would like to characterize.
We remark that the versatility of Eq.\ \eqref{F_C} opens the possibility to find criteria for other entanglement structures. For example, note that $k$-producibility is insensitive to the spatial distribution of entanglement: Suppose the $N$ spins are arranged on a chain and are genuinely $2$-party entangled (i.e., they are not fully separable). One might thus want to be able to distinguish between whether the first two spins are entangled or the first and last spin on the chain are entangled as it might be much less challenging to prepare the former case than the latter. This fine-grained form of multiparticle entanglement is captured by the notion of $k$-wide entanglement introduced in \cite{Woelk} and easily incorporated in our framework by defining $\mathcal{C}$ as the set of configurations where the spins are at most a distance $k$ apart.

\begin{figure}[t]
	\includegraphics[width=0.49\textwidth]{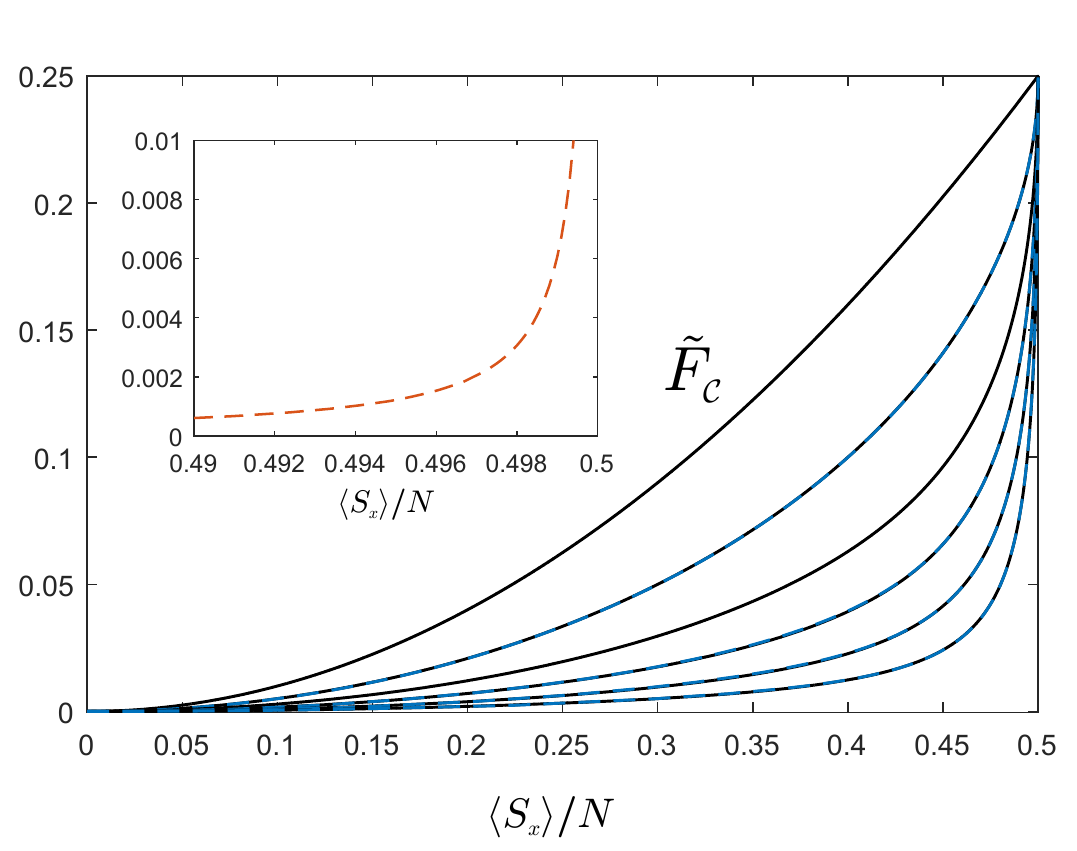}
	\caption{\label{figure}Detecting $k$-party entanglement: Lower bounds $\tilde{F}_{\mathcal{C}}$ to the functions $F_{\mathcal{C}}$ in Eq.~\eqref{k-local bounds} for the three coinciding cases $q=0,\pi/2,$ and $\pi$, with $S=1/2$, and $k = 1,2,5,10,20,40$ for any $N$ obtained from the SDP described in Appendix \ref{app:lower bound}. To compare, we also show the bounds from Ref.\ \cite{sorensen} for $k$ even (light blue, dashed). Under the assumption {(2)} as described in the main text the bounds can be calculated for large $k$ with numerical effort independent of $k$. For demonstration, we show the bound for $k = 2\cdot 10^4$ (inset).  If a measurement lies below a curve corresponding to $k$ in the plot then the state is $(k+1)$-party entangled.}
\end{figure}

\section{Non-permutationally symmetric extreme spin squeezing}
To give concrete examples, we now focus on spins arranged on a chain and consider the operators
\begin{equation}
\label{struct_factor}
\hat O=\frac{1}{\sqrt{N}}\hat S_{z}(q) = \frac{1}{\sqrt{N}}\sum_{j=1}^N \me^{\mi q j}\hat S_{z}^{(j)}
\end{equation}
and $\hat M=\hat S_{x}(0)/N=:\hat S_{x}/N$. Here, $\hat S_{z}^{(j)}$ denotes a spin-$S$ operator along the $z$ direction acting on the $j$'th spin and $q \in [0,2\pi]$ and we choose $\hat M$ to be the magnetization per spin in the direction $x$. 
As a consequence of our main observation it follows that for every state $\hat\varrho$ in the class $\mathcal{C}$ we have
\begin{equation}
\label{k-local bounds}
\frac{\Delta_{\hat \varrho}^2[\hat S_{z}(q)]}{N} \ge F_{\mathcal{C}}\left(\frac{\langle \hat S_{x}\rangle_{\hat\varrho}}{N}\right),\\
\end{equation}
with $F_{\mathcal{C}}$ as in Eq.\ \eqref{F_C}. 

Inequality \eqref{k-local bounds} includes the $k$-party bounds of S\o rensen and M\o lmer as a special case with $q=0$ and the corresponding set $\mathcal{C}$. The inequality of S\o rensen and M\o lmer is maximally violated by the ground state of the so-called spin-squeezing Hamiltonian, i.e.\ the one axis twisting plus external field, that has been shown to be an enhancement over the simple one-axis twisting. Experimentally, the bound has been able to confirm multiparticle entanglement in various setups \cite{gross,hosten,riedel,cox,engels,dellantonio}.

For this important special case, the SDP method presented in Appendix \ref{app:lower bound} for spin-1/2 gives a reliable lower bound to $F_{\mathcal{C}}$ that scales linearly with the size of the set $\mathcal{C}$ and, hence, linearly with $k$. Clearly, these bounds can then be used to obtain criteria for arbitrary spin-$S$ particles as well. Notably, besides the scalability, our approach avoids all assumptions the proof of the bounds in \cite{sorensen} has been relying on and, hence, makes technically subsequent publications \cite{Vitagliano2, luecke, gross, hosten, riedel, cox, engels, dellantonio} rigorous that refer to the original work.

\om{More specifically, {(1)} convexity of the bounds in \cite{sorensen} has been one of the requirements of the proof and which is verified numerically can either be investigated numerically or follows as a result of assumption {(3)} discussed below. From numerical inspections for small $k$ one may also infer that {(2)} the optimal configuration in Eq.\ \eqref{F_C} is $\mathcal{Z} = \{1,\ldots,k-1\}$ and, {(3)} for $|\mathcal{Z}|$ even, the infimum in Eq.\ \eqref{G_Z} (with the above mentioned observables) is achieved for $\langle \hat S_z \rangle_{\hat \varrho} = 0$. The latter assumption transforms the variance-minimization into a simple ground state search that can be solved efficiently, in particular, if {(4)} the ground state is assumed to lie in the symmetric subspace of dimension $k+1$. In contrast, the case $|\mathcal{Z}|$ odd remains numerically more costly \cite{sorensen, dellantonio} and is usually omitted. (1) can be also considered the direct consequence of (3) and the fact that the set of points corresponding to physical states in the $(\langle\hat S_x\rangle, \langle\hat S_z^2\rangle)$-space is convex.}

Importantly, our method {does not need these assumptions and there is no technical difference between an even and odd number of spins}. We can thus, indeed, consider the optimization over all configurations in Eq.\ \eqref{F_C} and, moreover, efficiently determine bounds for $k$ even and odd. We observe numerically very good agreement with the bounds obtained by previous methods and that they improve with increasing $k$, see Fig. \ref{figure}.

As noted before, not only for this special case but for any $q$, the observables that appear in Eq.\ \eqref{k-local bounds} are of practical importance since, e.g., the generalized variance may be accessible in scattering experiments \cite{hart,mazurenko}. Then, depending on the value of $q$, symmetries may simplify the different steps that are required to numerical compute $F_{\mathcal{C}}$ significantly. On one hand, the operators may exhibit an efficient parametrization, see Appendix \ref{symmetries} for a discussion of symmetries of the operators under consideration. On the other hand, in general, in order to obtain a criterion for $k$-party entanglement, the minimization over all subsets $\mathcal{Z}\subset [N]$ of cardinality at most $k$ needs to be considered. Here, the presence of symmetries, that may also be present for $q\ne 0$, reduces the number of ways to select $k$ out of $N$ spins that may lead to distinct bounds. For example, taking translational symmetries into account, the number of inequivalent subsets may be counted (see Appendix \ref{symmetries}) and determined numerically \cite{sawada}. To provide an example, for $q=\pi/2$, $S=1/2$, and $k$ up to $40$ we compute the lower bound $\tilde{F}_{\mathcal{C}}$ to $F_{\mathcal{C}}$ for $k$-producible  states (i.e., $\mathcal{C}$ as in Eq.~\eqref{k producible C}) using the SDP approach \cite{baumgratz} and Appendix \ref{app:lower bound}, see Fig.\ \ref{figure}. Note that the given bounds for $k$-local states are valid for any number of spins $N$, again a consequence of symmetries, see Appendix \ref{symmetries}.

\section{Engineering $k$-party Entangled States by a Quantum Quench in Ion Traps}
Modern experimental platforms such as trapped ions allow for the implementation of quantum systems with a wide range of tunable interactions. \om{This has raised the interest in control and study of spin systems with artificial interactions that, for example, can result in exotic quantum phases and novel quantum states \cite{hauke, Hauke2015, lin, chiuri} where the quantification of entanglement can help to characterize those quantum effects \cite{marty2, cramer2, daley, islam}.}
We find numerically and demonstrate below that states violating inequality Eq.\ \eqref{k-local bounds} can be generated under a time-evolution. In particular, one may consider a quantum quench under an Ising Hamiltonian with a transverse field
\begin{equation}
\label{ising hamiltonian}
\hat H = \sum_{i<j} J_{i,j} \hat\sigma_z^{i}\hat\sigma_z^{j}  + B \sum_i \hat\sigma_x^{i},
\end{equation}
with couplings of the form
\begin{equation}
\label{cos interactions}
J_{i,j} \propto \cos q(i-j).
\end{equation}
The protocol is described by initializing the system in a fully polarized state $\ket{\Psi_0} = \ket{\uparrow\uparrow \cdots \uparrow}_x$ parallel to the transverse magnetic field and let it evolve under the Hamiltonian in Eq.~\eqref{ising hamiltonian}.

\begin{figure}[t]
	\includegraphics[width=0.49\textwidth]{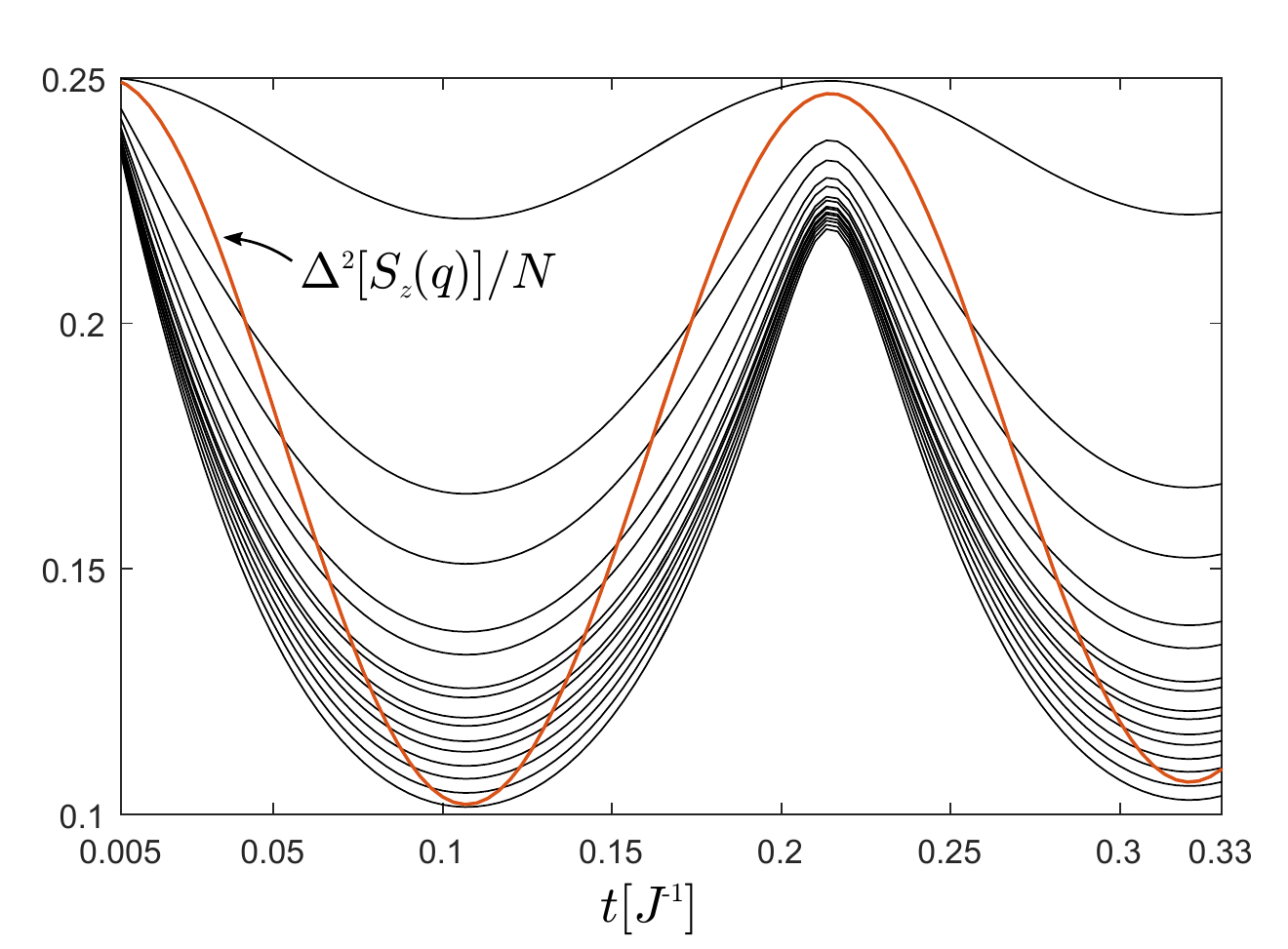}
	\caption{\label{figure2} Detection of $k$-party entanglement after a quench under the Hamiltonian in Eq.~\eqref{ising hamiltonian} with couplings as in Eq.~\eqref{cos interactions}. Shown as a function of time are $\Delta^2[\hat S_{z}(2\pi/16)]/N$ (red), and the lower bounds $F_{\mathcal{C}}$ evaluated at the instantaneous magnetization $\langle \hat S_x \rangle/N$ (black) for $N=15$ spins. At $t\approx 0.11 J^{-1}$ we observe the maximum of 15-party entanglement.}
\end{figure} 

Experimentally, for some specific values of $q$ and $N$ this may be achieved with trapped ions. Today's ion trap technologies allow for the implementation of a Hamiltonian of the form Eq.\ \eqref{ising hamiltonian} with interactions given by (see. e.g., \cite{Kim})
\begin{equation}
\label{effective interactions}
J_{i,j} \propto \sum_{\alpha} \Omega_i^{\alpha}\Omega_j^{\alpha} \sum_{n=1}^N\frac{(\vec{b}_{\alpha,n})_i(\vec{b}_{\alpha,n})_j}{\mu_{\alpha}^2 - (\omega_n^{\alpha})^2},
\end{equation}
where $\vec{b}_{\alpha,n}$ denotes the eigenvector corresonding to the $n$'th eigenmode of the system, $\Omega_i^{\alpha}$ the Rabi frequency on the $i$'th ion,  $\mu_{\alpha}$ the laser detuning and $\omega_n^{\alpha}$ the frequency of the $n$'th eigenmode. These quantities may depend on the direction $\alpha = x,y,z$.
In a trapped-ion system where the couplings are effectively described by Eq.\ \eqref{effective interactions} one can use the freedom of controlling the Rabi frequencies $\Omega_i^{\alpha}>0$ and detuning $\mu_{\alpha}$ in order to generate interactions of the form Eq.\ \eqref{cos interactions}. The crucial observation is that the coupling matrix Eq.\ \eqref{cos interactions} has rank two with eigenvectors that may resemble two of the transversal eigenmodes $\vec{b}_{\alpha,n}$. By resonantly addressing these two modes separately from the two transversal directions and by an adjustment of the Rabi frequencies one obtains interactions as in Eq.~\eqref{cos interactions}. This is possible for specific values of $q$ as described in Appendix \ref{engineering the coupling}.

We observe that, for the above described quench protocol, the entanglement of the system can be detected by the criterion in Eq.\ \eqref{k-local bounds} as shown in Fig.\ \ref{figure2}. \om{The bounds certify that the spins become genuine multipartite entangled}. Note that one may also prepare the ground state of the Hamiltonian Eq.\ \eqref{ising hamiltonian} instead. However, preparing a state with a quench may experimentally be easier to accomplish than the ground state, e.g., via an adiabatic ramp, due to its shorter duration.

\section{Summary and Outlook}

In conclusion, we have introduced a method to derive criteria for the detection of various many-body entanglement structures, with emphasis on $k$-party entanglement. Other entanglement structures such as $k$-partite or $k$-wide \cite{Woelk} entanglement are immediately covered by our scheme if an additional optimization over specific spin configurations is taken into account. The criteria make no assumptions on the state, require to measure a few global observables only and are applicable to any number of total spins. In contrast to previous works, the observables do not have to be permutation invariant. Instead, even if there is no symmetry identified, we show how to compute the multiparty entanglement criteria, where the algorithmic approach we take can be further extended using DMRG which may allow for the investigation of systems of many spins. We leave this exploration to future work. We find that for the case of permutation invariant observables, our approach enables to derive the bounds without any technical assumption. As an application of the method we provide an experimental protocol to test the entanglement criteria under realistic conditions with trapped ions.

\section{Acknowledgements}

We acknowledge the support of the ERC synergy grant BioQ, the EU project QUCHIP, the EU (ERC Starting Grant 258647/GEDENTQOPT,  CHIST-ERA QUASAR, COST Action CA15220), the Spanish Ministry of Economy, Industry and Competitiveness and the European Regional Development Fund FEDER through Grant No.\ FIS2015-67161-P (MINECO/FEDER), the Basque Government (Project No.\ IT986-16), the UPV/EHU program UFI 11/55 and the Austrian Science Fund (FWF) through the START project Y879-N27. This work was supported by the DFG through SFB 1227 (DQ-mat) and the RTG 1991, the ERC grants QFTCMPS and SIQS, and the cluster of excellence EXC201 Quantum Engineering and Space-Time Research. The authors acknowledge support by the state of Baden-W\"urttemberg through bwHPC.




\appendix

\onecolumngrid

\section{Main observation}
\label{app:main observation}
For any state $\varrho$ and any operator $O$ we have
\begin{equation}
\begin{split}
\Delta_\varrho^2[O]&=\langle O^\dagger O\rangle_\varrho-\langle O^\dagger\rangle_\varrho\langle  O\rangle_\varrho\\
&=\langle O^\dagger O\rangle_\varrho-\langle O\rangle_\varrho^*\langle  O\rangle_\varrho+\inf_{s\in\cc}\left|s-\langle O\rangle_\varrho\right|^2\\
&=\inf_{s\in\cc}\left[\langle O^\dagger O\rangle_\varrho+|s|^2-s^*\langle O\rangle_\varrho-s\langle O\rangle_\varrho^*\right]\\
&=\inf_{s\in\cc}\langle (O-s\id)^\dagger (O-s\id)\rangle_\varrho
\end{split}
\end{equation}
and (as the second line shows) the minimum is attained at $s=\langle  O\rangle_\varrho$. Using this variational form of the generalized variance, we find that for any state $\varrho$, any operator $O$, any $\lambda\in\rr$, and any Hermitian operator $M$,
\begin{equation}
\begin{split}
\Delta_\varrho^2[O]&=\lambda\langle M\rangle_\varrho+\Delta_\varrho^2[O]-\lambda\langle M\rangle_\varrho\\
&=\lambda\langle M\rangle_\varrho+\inf_{s\in\cc}\langle \left[ (O-s\id)^\dagger (O-s\id)-\lambda M\right]\rangle_\varrho\\
&\ge\lambda\langle M\rangle_\varrho+\inf_{s\in\cc}\lambda_{\text{min}} \left[ (O-s\id)^\dagger (O-s\id)-\lambda M\right]\\
&=:\lambda\langle M\rangle_\varrho+G_{[N]}(\lambda),
\end{split}
\end{equation}
where $\lambda_{\text{min}}[\cdot]$ denotes the smallest eigenvalue of the Hermitian matrix in brackets. Note that $\lambda_{\text{min}} \left[ (O-s\id)^\dagger (O-s\id)-\lambda M\right]$ is concave in $\lambda$ such that $G_{[N]}(\lambda)$ is also concave in $\lambda$.

Suppose now that $\varrho$ is of the product form 
\begin{equation}
\label{prod_state_P}
\varrho_1\otimes\cdots\otimes \varrho_{P}.
\end{equation}
This divides the $N$ spins into sets of spins $\mathcal{Z}_p \subset \{1,\ldots,N\}$, $p = 1,\ldots,P$, which form a partition $\bigcup_p \mathcal{Z}_p = \{1,\dots,N\}$ and $\mathcal{Z}_p$ denotes the set of spins that $\varrho_p$ acts on. If we further assume that
$O=\sum_{i=1}^NO_i$ and $M=\sum_{i=1}^NM_i$ then (we use the shorthand notation $O_{\mathcal{Z}}=\sum_{i\in\mathcal{Z}}O_i$ and similarly for $M$), we find for states as in Eq.~\eqref{prod_state_P}
\begin{equation}
\begin{split}
\Delta_\varrho^2[O]&=\sum_{p=1}^P\Delta_{\varrho_p}^2[O_{\mathcal{Z}_p}]\\
&\ge \sum_{p=1}^P\left(\lambda\langle M_{\mathcal{Z}_p}\rangle_{\varrho_p}+|\mathcal{Z}_p|\frac{G_{\mathcal{Z}_p}(\lambda)}{|\mathcal{Z}_p|}\right)\\
&= \lambda\langle M\rangle_{\varrho}+\sum_{p=1}^P |\mathcal{Z}_p|\frac{G_{\mathcal{Z}_p}(\lambda)}{|\mathcal{Z}_p|}.
\end{split}
\end{equation}
Now suppose that all the $\mathcal{Z}_p$ are elements of some set $\mathcal{C}$. Then for all states $\varrho$ as in Eq.~\eqref{prod_state_P}
\begin{equation}
\begin{split}
\Delta_\varrho^2[O]&\ge\lambda\langle M\rangle_\varrho+\sum_{p=1}^P|\mathcal{Z}_p|\min_{\mathcal{Z}\in\mathcal{C}}\frac{G_{\mathcal{Z}}(\lambda)}{|\mathcal{Z}|}\\
&=\lambda\langle M\rangle_\varrho+N\min_{\mathcal{Z}\in\mathcal{C}}\frac{G_{\mathcal{Z}}(\lambda)}{|\mathcal{Z}|}.
\end{split}
\end{equation}
Finally, we may extend the above to states that are convex combinations of states as in Eq.~\eqref{prod_state_P}, i.e., for states of the form 
\begin{equation}
\varrho=\sum_np_n\varrho^{(n)}_1\otimes\cdots\otimes \varrho^{(n)}_{P_n}=:\sum_np_n\varrho_n,\;\;\; p_n\ge 0,\;\;\;\sum_np_n=1, 
\end{equation}
by the Cauchy--Schwarz inequality:
\begin{equation}
\begin{split}
\Delta_\varrho^2[O]&=\sum_np_n\langle O^\dagger O\rangle_{\varrho_n}-\sum_{n,m}p_np_m\langle O^\dagger\rangle_{\varrho_n}\langle  O\rangle_{\varrho_m}\\
&\ge \sum_np_n\langle O^\dagger O\rangle_{\varrho_n}-\sum_{n}p^2_n|\langle O\rangle_{\varrho_n}|^2\\
&\ge \sum_np_n\Delta_{\varrho_n}^2[O]\\
&\ge \lambda\langle M\rangle_{\varrho}+N\min_{\mathcal{Z}\in\mathcal{C}}\frac{G_{\mathcal{Z}}(\lambda)}{|\mathcal{Z}|}.
\end{split}
\end{equation}
As this holds for all $\lambda\in\rr$, we may take the supremum to arrive at
\begin{equation}
\begin{split}
\Delta_\varrho^2[O]&\ge F_{\mathcal{C}}(\langle M\rangle_\varrho),\;\;\;F_{\mathcal{C}}(m)=\sup_{\lambda\in\rr}\left(\lambda m+N\min_{\mathcal{Z}\in\mathcal{C}}\frac{G_{\mathcal{Z}}(\lambda)}{|\mathcal{Z}|}\right),
\end{split}
\end{equation}
where we recall that $G_{\mathcal{Z}}$ is concave and note that $F_{\mathcal{C}}$ convex.

\section{Symmetries for $q=2\pi/z$}

\label{symmetries}

In this section, we show how to exploit the symmetries of the operator Eq.\ \eqref{struct_factor} for $q=2\pi/z$ and $z$ integer in order to reduce the numerical effort to derive the bound $F_{\mathcal{C}}$ in Eq.\ \eqref{k-local bounds}. By the periodicity of the phases, for some set $\mathcal{Z} \subset [N]$, we may write
\begin{equation}
\begin{split}
\hat O_{\mathcal{Z}} &= \frac{1}{\sqrt{N}}\sum_{j\in\mathcal{Z}} \me^{\mi q j} \hat S_z^{(j)} \\
&=: \frac{1}{\sqrt{N}}\sum_{n = 1}^{\zeta}\me^{\mi q n}\sum_{j\in\mathcal{Z}^n}  \hat S_z^{(j)},
\end{split}
\end{equation}
where, since $F_{\mathcal{C}}$ in Eq.\ \eqref{k-local bounds} (see also the definitions Eq.\ \eqref{F_C} and \eqref{G_Z}) is invariant under local spin flips in $z$-direction, we may consider $\zeta=z$ for $z$ odd and $\zeta=z/2$ for $z$ even, and define $\mathcal{Z}^n := \left\{j \in \mathcal{Z}\; \big|\; (\me^{\mi q j} = \me^{\mi q n}) \vee (\me^{\mi q j} = -\me^{\mi q n}) \right\} \subset \mathcal{Z}$.

Therefore, the bounds for $z=1$ coincides with the bounds for $z=2$, since in both cases $\zeta = 1$. For $z=4$, i.e., $\zeta = 2$,
\begin{equation}
\hat O_{\mathcal{Z}} = \frac{1}{\sqrt{N}} \left( \mi \sum_{j\in\mathcal{Z}^1}  \hat S_z^{(j)} + \sum_{j\in\mathcal{Z}^2}  \hat S_z^{(j)} \right).
\end{equation}
With this, we find that

\begin{equation}
\begin{split}
G_{\mathcal{Z}}(\lambda) &= \inf_{s\in \cc} \lambda_{\min}\left[(\hat O_{\mathcal{Z}}(\vec{j}) - s \id)^\dagger(\hat O_{\mathcal{Z}}(\vec{j}) - s \id) - \lambda \hat M_{\mathcal{Z}}(\vec{j})\right], \\
&= \inf_{s\in \cc} \lambda_{\min}\left[ \left(\frac{1}{\sqrt{N}}\sum_{j\in\mathcal{Z}^1} \hat S_z^{(j)} - \Im (s) \right)^\dag \left(\frac{1}{\sqrt{N}}\sum_{j\in\mathcal{Z}^1} \hat S_z^{(j)} - \Im (s) \right) - \frac{\lambda}{N} \sum_{j\in\mathcal{Z}^1} \hat S_x^{(j)} \right. \\
& \hspace{55pt} \left. + \left(\frac{1}{\sqrt{N}}\sum_{j\in\mathcal{Z}^2} \hat S_z^{(j)} - \Re (s) \right)^\dag \left(\frac{1}{\sqrt{N}}\sum_{j\in\mathcal{Z}^2} \hat S_z^{(j)} - \Re (s) \right) - \frac{\lambda}{N} \sum_{j\in\mathcal{Z}^2} \hat S_x^{(j)} \right] \\
&= \inf_{\Im (s)\in \rr} \lambda_{\min}\left[ \left(\frac{1}{\sqrt{N}}\sum_{j\in\mathcal{Z}^1} \hat S_z^{(j)} - \Im (s) \right)^\dag \left(\frac{1}{\sqrt{N}}\sum_{j\in\mathcal{Z}^1} \hat S_z^{(j)} - \Im (s) \right) - \frac{\lambda}{N} \sum_{j\in\mathcal{Z}^1} \hat S_x^{(j)} \right] \\
& + \inf_{\Re (s)\in \rr} \lambda_{\min}\left[\left(\frac{1}{\sqrt{N}}\sum_{j\in\mathcal{Z}^2} \hat S_z^{(j)} - \Re (s) \right)^\dag \left(\frac{1}{\sqrt{N}}\sum_{j\in\mathcal{Z}^2} \hat S_z^{(j)} - \Re (s) \right) - \frac{\lambda}{N} \sum_{j\in\mathcal{Z}^2} \hat S_x^{(j)} \right]. \\
\end{split} 
\end{equation}

That is, the case $z = 4$ can also be expressed in terms of the permutation invariant case.

Now, note that for all $n$ the operators $\hat O_{\mathcal{Z}}$ and $\hat M_{\mathcal{Z}} = \sum_{j\in \mathcal{Z}} \hat S_x^{(j)}$ are invariant under permutations of the sites in $\mathcal{Z}^n$. Any operator with these symmetry properties may be represented by an operator $\hat X$ with a decomposition as (see e.g.\ Ref.\ \cite{novo})
\begin{equation}
\label{PIblockstructure}
\hat X = \bigoplus_{\vec{j} = (j_1,\ldots,j_{\zeta})} \sum_{m} \hat X_m^{[1]}(j_1) \otimes \cdots \otimes \hat X_{m}^{[\zeta]}(j_{\zeta}) =: \bigoplus_{\vec{j}} \hat X (\vec{j}),
\end{equation}
where $j_n = j_n^{\min},j_n^{\min} + 1, \ldots, |\mathcal{Z}^n|/2$ with $j_n^{\min} = 0$ $(j_n^{\min} = 1/2)$ if $|\mathcal{Z}^n|$ is even (odd) and where $\hat X_{m}^{[n]}(j_n)$ are of dimension $(2j_n + 1) \times (2j_n + 1)$. Now, every eigenvector of $\hat X$ belongs to one of the blocks in the decomposition Eq.\ \eqref{PIblockstructure}. To calculate $G_{\mathcal{Z}}(\lambda)$ we may therefore minimize each block separately, and obtain

\begin{equation}
G_{\mathcal{Z}}(\lambda) = \min_{\vec{j}} \inf_{s\in \cc} \lambda_{\min}\left[(\hat O_{\mathcal{Z}}(\vec{j}) - s \id)^\dagger(\hat O_{\mathcal{Z}}(\vec{j}) - s \id) - \lambda \hat M_{\mathcal{Z}}(\vec{j})\right],
\end{equation}

where $\hat O_{\mathcal{Z}}(\vec{j}) = \sum_{n=1}^{\zeta} \me^{\mi q n} \hat S_z(j_n)$ and $M_{\mathcal{Z}}(\vec{j}) = \sum_{n=1}^{\zeta} \hat S_x(j_n)$, with $\hat S_{\alpha}(j)$ the $\alpha$-component spin-$j$ operator.

Notably, for every configuration of spins $\mathcal{Z}$, $G_{\mathcal{Z}}$ depends the cardinality of the subsets $|\mathcal{Z}^n|$ only. Moreover, the variance of $\hat O_{\mathcal{Z}}$ in Eq.\ \eqref{PIblockstructure} only depends on $\me^{\mi q (m-n)}$, for $m,n = 1,\ldots,\zeta$. Therefore we may consider soley configurations with cardinalities $|\mathcal{Z}^n|$ inequivalent under cyclic shifts and reflection of the index $n$, i.e.\ inequivalent under $n \mapsto n + l$ for $l \in \nn$, and $n \mapsto \zeta - n + 1$. Only configurations that cannot be obtained from one another by these operations will result in different bounds.

Now, in order to derive the bound for $k$-party entanglement we need to consider all inequivalent configurations $\mathcal{Z}$ with cardinality at most $k$. 
This then gives the size of the set $\mathcal{C}$ to obtain bounds for multiparty entanglement for the observables of consideration. In general, this may be done numerically. For the special case where $N$ is divisible by $\zeta$ we count the number of these configurations for  for fixed $k$ using P\'olya's enumeration theorem (PET) \cite{redfield, polya}. We set out to count the number of ways one can assign a number $0 \le k(n) \le k_{\max}$, where $k_{\max} := N/\zeta$, to every phase labelled by $n = 1,\ldots,\zeta$ with the constraint $\sum_{n=1}^\zeta k(n) = N$. For a particular configuration or `coloring' $k:\{1,\ldots,\zeta\} \rightarrow \{0,\ldots,k_{\max}\}$, $k(n)$ determines the number of spins with phase factor $n$. As already noted, since only quantities of the form $\me^{\mi q (m-n)}$ are relevant, combinations that are equal up to a cyclic shifts and reflection will result in the same bound. Mathematically, two colorings $k$ and $k'$ are equivalent if there is a permutation $\tau$ of the set $\{1,\ldots,\zeta\}$ that belongs to the dihedral group $D_\zeta$ and is such that $k' = k\circ \tau^{-1}$.
We introduce the generating function, a polynomial in $k_{\max}$ variables $F_{D_\zeta}(r_0,\ldots,r_{k_{\max}}) = \sum_{p_0,\ldots,p_{k_{\max}}} f_{D_\zeta}(p_0,\ldots,p_{\max}) r_0^{p_0}\cdots r_{k_{\max}}^{p_{k_{\max}}}$ where $f_{D_\zeta}$ is the number of orbits, i.e.\ distinct configuartions, under $D_\zeta$ with fixed content. Hence, in order to count all orbits for a constant number of sites, we are interested in the coefficient sum
\begin{equation}
\label{eq coeff sum}
n_{k,N,\zeta} = \sum_{\sum n p_n = N} f_{D_\zeta}(p_0,\ldots, p_{k_{\max}}).
\end{equation}
By the PET
\begin{equation}
F_{D_\zeta}(r_1,\ldots, r_{p_{\max}}) = Z_{D_\zeta}\left(\sum_{n=0}^{k_{\max}} r_n,\ldots,\sum_{n=0}^{k_{\max}} r_n^\zeta\right),
\end{equation}
where $Z_{D_{\zeta}}$ is the so-called cycle index polynomial of the dihedral group, here for two colors, given by

\begin{equation}
Z_{D_{\zeta}}(t_1,t_2,\ldots) = \begin{cases} \frac{1}{2{\zeta}} \left( \sum_{d|\zeta} \varphi(d) t_d^{\zeta/d} + \zeta t_1 t_2^{(\zeta-1)/2} \right), & \text{$\zeta$ odd,}\\ \frac{1}{2{\zeta}} \left( \sum_{d|z} \varphi(d) t_d^{\zeta/d} + \frac{\zeta}{2} t_1^2 t_2^{(\zeta-2)/2} + \frac{\zeta}{2} t_2^{\zeta/2} \right), & \text{$\zeta$ even,} \end{cases} 
\end{equation}
where the sum runs over all divisors $d$ of $\zeta$ and $\varphi$ denotes Euler's totient function.
Hence, we may use the cycle index to calculate Eq.\ \eqref{eq coeff sum}. We find
\begin{equation}
\label{number configurations}
n_{k,N,\zeta} = \frac{1}{2\zeta} \sum_{d|\textrm{gcd}(k,\zeta)} \varphi(d) \sum_{\vec{p} \in S_1(d)} {\zeta/d \choose \vec{p}} + \begin{cases} \frac{1}{2} \sum_{n=0}^{k_{\max}} \sum_{\vec{p} \in S_2(n)}{(\zeta-1)/2 \choose \vec{p}}, & \text{$\zeta$ odd,}\\ \frac{1}{4} \left(\sum_{\vec{p} \in S_1(2)} {\zeta/2 \choose \vec{p}} + \sum_{(\vec{p},\tilde{\vec{p}})\in S_3} {2 \choose \vec{p}} {(\zeta-2)/2 \choose \tilde{\vec{p}}} \right), & \text{$z$ even,} \end{cases}
\end{equation}
where
\begin{equation}
\begin{split}
S_1(d) &= \left\{(p_0,\ldots,p_{k_{\max}}) \Bigg| {\sum_{n=0}^{k_{\max}} p_n = \zeta/d,\; \sum_{n=1}^{k_{\max}} n p_n = k/d}\right\}, \\
S_2(m) &= \left\{(p_0,\ldots,p_{k_{\max}}) \Bigg| {\sum_{n=0}^{k_{\max}} p_n = (\zeta-1)/2,\; 2 \sum_{n=1}^{k_{\max}} n p_n = k-m}\right\}, \\
S_3 &= \left\{(p_0,\ldots,p_{k_{\max}},\tilde{p}_0,\ldots,\tilde{p}_{k_{\max}}) \Bigg|\sum_{n=0}^{k_{\max}} p_n = 2,\; \sum_{n=0}^{k_{\max}} \tilde{p}_n = (\zeta-2)/2,\; \sum_{n=1}^{k_{\max}} n (p_n + 2 \tilde{p}_n) = k \right\}.
\end{split}
\end{equation}

\begin{figure*}[t]
	\includegraphics[width=1\textwidth]{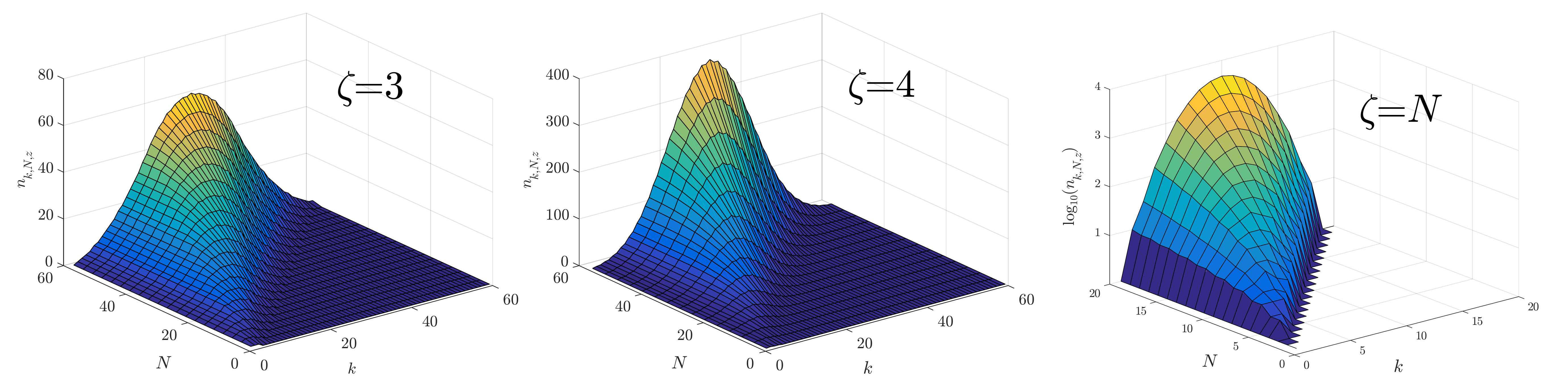}
	\caption{\label{figurepolya} The function $n_{k,N,\zeta}$ Eq.\ \eqref{number configurations} which gives the number of ways choosing $k$ spins out of $N$ that are inequivalent under the dihedral group $D_z$. This gives an upper bound on the number of configuration one has to optimize over in order to find a bound for $k$-party bound from observables of the form Eq.\ \eqref{struct_factor} for $\zeta$ different phases.}
\end{figure*}

\section{About the algorithm}
\label{Algorithm}

Here, we describe an algorithm to solve numerically the optimization problem of Eq. \eqref{G_Z}. More specifically, we need to solve a global eigenvalue minimization, i.e.\ to minimize the lowest eigenvalue, $\lambda_{\min}(\vec{x}) =\lambda_{\min}(\mathcal{A}(\vec{x}))$ of a matrix-valued function
\begin{equation}
\label{operator valued function}
\mathcal{A}(\vec{x}) = (O-s\id)^\dagger(O-s\id) + \lambda M,
\end{equation}
over a box $\mathcal{B} \subset \rr^2$ given by the conditions $x_1 \in [\lambda_{\min}[(O^\dagger + O)/2],\lambda_{\max}[(O^\dagger + O)/2]]$ and $x_2 \in [\lambda_{\min}[\mi(O^\dagger - O)/2],\lambda_{\max}[\mi(O^\dagger - O)/2]]$ and $s = x_1 + \mi x_2$.
Generally speaking, the difficulty of the optimization problem comes from its nonconvexity. The algorithm described in \cite{mengi} adresses this challenge by introducing quadratic support functions that provide a lower bound to the eigenvalue function $\lambda_{\min}(\vec{x})$.  The determination of the support functions relies on a global lower bound $\gamma$ to $\lambda_{\min}[\nabla^2\lambda_{\min}(\vec{x})]$ which requires the analyticity of the eigenvalue function. For any $\vec{x}_0 \in \mathcal{B}$ where $\Lambda(\vec{x})$ is non-degenerate, a support function is given by
\begin{equation}
\label{quadratic support}
q(\vec{x}) = \lambda_{\min}(\vec{x}_0) + \nabla\lambda_{\min}(\vec{x}_0)^T(\vec{x} - \vec{x}_0) + \frac{\gamma}{2} \|\vec{x} - \vec{x}_0\|^2.
\end{equation}
Thus, to determine the support function we need to evaluate $\lambda_{\min}(\vec{x}_0)$ and the gradient $\lambda_{\min}(\vec{x}_0)$, which is given by
\begin{equation}
\label{gradient}
(\nabla \lambda_{\min}(\vec{x}))_j = \bra{\Psi_0}\frac{\partial \mathcal{A}(\vec{x})}{\partial x_j}\ket{\Psi_0},
\end{equation}
where $\ket{\Psi_0}$ denotes the eigenstate of $\mathcal{A}$ to the lowest eigenvalue $\lambda_{\min}(\vec{x})$. 

\begin{figure*}[t]
	\includegraphics[width=0.98\textwidth]{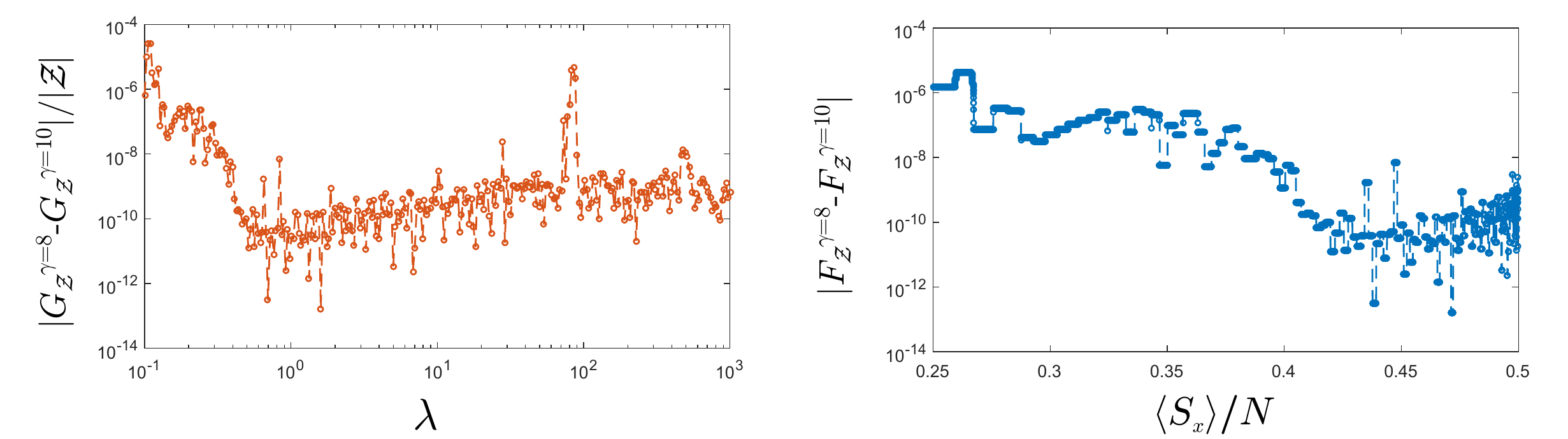}
	\caption{\label{figure4} As an example we consider the error between the bound Eq.\ \eqref{k-local bounds} $F_{\mathcal{C}}$ to detect $(k+1)$-party entanglement with $k=15$ and $q=2\pi/16$ obtained for $\gamma=8$ and $\gamma = 10$. For a large magnetic field which corresponds to an expectation value of $\langle S_x \rangle / N$ close to 0.5 the problem is convex, as may be seen from direct inspection whereas for smaller external fields the problem becomes nonconvex (not shown). In the example, the algorithm stops either when the gap between the support function Eq.\ \eqref{quadratic support} and the function value becomes smaller than $10^{-8}$ or when a maximum of 1500 iteration is reached.}
\end{figure*}

For large systems, we can use DMRG to determine $\lambda_{\min}(\vec{x})$ as well as $\ket{\Psi_0}$ and calculate the gradient Eq.\ \eqref{gradient} exploiting the fact that expectation values of matrix-product operators with matrix-product states can be determined efficiently.
To study the second derivative of $\lambda_{\min}(\vec{x})$ in more details, we assume that $\lambda_{\min}(\vec{x})$ is non-degenerate for all $\vec{x}$ inside the parameter range defined above. The Hessian of $\lambda_{\min}(\vec{x})$ is then given by (see section 3.2.3 of \cite{mengi})

\begin{equation}
(\nabla^2 \lambda_{\min}(\vec{x}))_{i,j} = \bra{\Psi_0} \frac{\partial^2 \mathcal{A}}{\partial x_i \partial x_j}\ket{\Psi_0} - 2 \sum_{k>0} \frac{1}{\lambda_k(\vec{x}) - \lambda_0(\vec{x})} \mathrm{Re}\left[\bra{\Psi_0}\frac{\partial \mathcal{A}}{\partial x_i} \ketbra{\Psi_k}{\Psi_k} \frac{\partial \mathcal{A}}{\partial x_j} \ket{\Psi_0} \right],
\end{equation}

where $\lambda_k$ and $\ket{\Psi_k}$ denote the $k$th smallest eigenvalue and corresponding eigenvector, respectively, of $\mathcal{A}$. Moreover,
\begin{equation}
\begin{split}
\frac{\partial^2 \mathcal{A}}{\partial x_i \partial x_j} &= 2 \delta_{i,j} \id, \\
\frac{\partial \mathcal{A}}{\partial x_1} &= -(O^\dagger + O) + 2 \id x_1, \\
\frac{\partial \mathcal{A}}{\partial x_2} &= -\mi(O^\dagger - O) + 2\id x_2.
\end{split}
\end{equation}
In particular for $O = S_{\alpha}(q)$ and $M = S_{\beta}$ one obtains
\begin{equation}
\label{structurefactor partial}
\begin{split}
\frac{\partial \mathcal{A}}{\partial x_1} &= 2\left(\id-\sum_{j=1}^N \cos(qj) S_{\alpha}^{j}\right), \\
\frac{\partial \mathcal{A}}{\partial x_2} &= 2\left(\id+\sum_{j=1}^N \sin(qj) S_{\alpha}^{j}\right).
\end{split}
\end{equation}
A lower bound to the minimal eigenvalue of $\nabla^2 \Lambda$ in terms of the spectral gap may be given as

\begin{equation}
\begin{split}
\lambda_{\min}[\nabla^2 \lambda_{\min}(\vec{x})] &= 2 - 2 \lambda_{\max}\left( \left[\sum_{k>0} \frac{1}{\lambda_k(\vec{x}) - \lambda_0(\vec{x})} \mathrm{Re}\left[\bra{\Psi_0}\frac{\partial \mathcal{A}}{\partial x_i} \ketbra{\Psi_k}{\Psi_k} \frac{\partial \mathcal{A}}{\partial x_j} \ket{\Psi_0} \right]\right]_{i,j} \right) \\
&\ge 2\left(1 - \frac{\lambda_{\max}\left[ \mathrm{Re}[C] \right]}{\lambda_1(\vec{x}) - \lambda_0(\vec{x})}\right),
\end{split}
\end{equation}

where
\begin{equation}
C_{i,j} = \bra{\Psi_0}\frac{\partial \mathcal{A}}{\partial x_i} \frac{\partial \mathcal{A}}{\partial x_j} \ket{\Psi_0} -\bra{\Psi_0}\frac{\partial \mathcal{A}}{\partial x_i} \ketbra{\Psi_0}{\Psi_0} \frac{\partial \mathcal{A}}{\partial x_j} \ket{\Psi_0}.
\end{equation}
Hence, whenever $\mathcal{A}$ has a non-degenerate ground state for all $x \in \mathcal{B}$, there is a global lower bound to $\lambda_{\min}[\nabla^2 \lambda_{\min}]$. In the numerical examples shown in Fig.\ \ref{figure2} we increase $\gamma$ until we do not observe any change in the bound. as shown in Fig.\ \ref{figure4}.

\section{Engineering the couplings in ion traps}
\label{engineering the coupling}
For completeness, we start this section by summarizing the derivation of the effective spin couplings as they can be generated with trapped ions. We consider the ions to be confined in a linear trap with the interactions generated by one bichromatic laser field for each of the two transversal directions $\alpha = x, y$ at frequencies $\omega_{\mathrm{atom}} \pm \mu_{\alpha}$, respectively. Here, $\omega_{\mathrm{atom}}$ denotes the level splitting of the internal two-level system used to encode the spin, e.g.\ the hyperfine clock states of an Ytterbium ion. The desired laser field can be achieved by two Raman beams per direction with corresponding frequency differences \cite{Kim}. The basic interaction of the $i$'th ion with a laser field at frequency $\omega$ and wave-vector $\vec{k}$ is given by (assume $\Omega \ge 0$)
\begin{equation}
H = \hbar\Omega \cos(\vec{k} \vec{\delta x} + \varphi + \omega t)\sigma_x^{i},
\end{equation}
where $\vec{\delta x}$ denotes the deviation of the ion from its equilibrium position.
Therefore the resulting interaction for the above considered laser fields with the ion chain is described by Hamiltonian
\begin{equation}
H = \hbar\sum_{\alpha = x,y}\sum_{i = 1}^N \Omega_i^{\alpha} \cos(\vec{\delta k}_{\alpha} \vec{\delta x}_i + \mu_{\alpha} t) \sigma_x^{i},
\end{equation}

where $(\vec{\delta k}_{\alpha})_{\xi} = \delta_{\xi,\alpha} \delta k$ is the wave-vector difference of the Raman beams in the direction $\alpha$ and $\vec{\delta x}_i$ denotes the deviation of the $i$'th ion from its equilibrium position. We may write $\delta k(\vec{\delta x}_i)_\alpha = \sum_{n=1}^N \eta_{i,n}^{\alpha} (a_{\alpha,n} + a_{\alpha,n}^{\dagger})$ with the Lamb-Dicke parameter $\eta_{i,n}^{\alpha} = \delta k (\vec{b}_{\alpha,n})_i\sqrt{\hbar/2M\omega_n^{\alpha}}$, where $M$ denotes the mass of the ions. Thereby, $\vec{b}_{\alpha,n}$ denotes the eigenvector of the $n$'th eigenmode in direction $\alpha$ and $a_{\alpha,n} (a_{\alpha,n}^{\dagger})$ the corresponding annihilation (creation) operators. Within Lamb-Dicke regime where $\eta_{i,n}^{\alpha}$ is small such that the condition $|\vec{\delta k}_{\alpha} \vec{\delta x}_i| \ll 1$ holds, and under the rotating-wave approximation justified by the condition $\omega_{\mathrm{atom}} \gg \mu_{\alpha} \gg \Omega_i^{\alpha}$, for $\alpha = x,y$ and $i = 1,\ldots, N$, one finds a state dependent spin-spin interaction mediated by the transversal eigenmodes of the trap. If furthermore $|\omega_n^{\alpha} - \mu_{\alpha}| \gg \eta_{i,n}^{\alpha} \Omega_i^{\alpha}$, also called the ``slow'' regime, excitation of the vibrational modes are only virtually excited and one obtains a spin Hamiltonian of the form $\sum_{i,j=1}^N J_{i,j} \sigma_x^{i}\sigma_x^{j}$ with \cite{Kim}
\begin{equation}
\label{effective couplings}
J_{i,j} =  \frac{(\hbar\delta k)^2}{2M}\sum_{\alpha = x,y} \Omega_i^{\alpha}\Omega_j^{\alpha} \sum_{n=1}^N\frac{(\vec{b}_{\alpha,n})_i(\vec{b}_{\alpha,n})_j}{\mu_{\alpha}^2 - (\omega_n^{\alpha})^2}.
\end{equation}

Next, we outline how Ising couplings of the form $J_{i,j} \propto \cos q(i-j)$, with $q = 2\pi/z$ and $z$ integer, may be designed with trapped ions. Since for all $i,j$ we have $\cos q(i-j) = \cos(qi + \varphi) \cos(qj + \varphi) + \sin(qi + \varphi) \sin(qj + \varphi)$ for any $\varphi$, and hence we may write
\begin{equation}
\label{cos couplings}
J \propto \lambda_c \vec{v}_c \vec{v}_c^T + \lambda_s \vec{v}_s \vec{v}_s^T,
\end{equation}
where $(\vec{v}_c)_i = \cos(qi + \varphi_N)$ and $(\vec{v}_s)_j = \sin(qj + \varphi_N)$ where $\varphi_N = \pi(1-(N+1)/z)$. Note that $\vec{v}_c$ and $\vec{v}_c$ are orthogonal and therefore proportional to the eigenvectors of $J$. On the other hand, the form of the matrix $(J_{i,j}) = J$ defined by the effective couplings Eq.\ \eqref{effective couplings} is mathematically equivalent to
\begin{equation}
J = \frac{(\hbar\delta k)^2}{2M} \sum_{\alpha,n} \vec{\beta}_{\alpha,n}\vec{\beta}_{\alpha,n}^T
\end{equation}
where $(\vec{\beta}_{\alpha,n})_i := \Omega_i^{\alpha}(\vec{b}_{\alpha,n})_i/\sqrt{\mu_{\alpha}^2 - (\omega_n^{\alpha})^2}$ for $\alpha = x,y$ and $i,n = 1,\ldots, N$. Suppose that we nearly resonantly excite two transversal modes, one in each of two directions $x$ and $y$,  which we denote by $m_x$ and $m_y$, respectively. The coupling matrix is approximately described by the matrix $J = \frac{(\hbar\delta k)^2}{2M} \left( \vec{\beta}_{x,m_x}\vec{\beta}_{x,m_x}^T + \vec{\beta}_{y,m_y}\vec{\beta}_{y,m_y}^T \right)$ and is hence of rank two. In order to mimick the couplings Eq.\ \eqref{cos couplings} for certain values of $q$ we may choose the Rabi frequencies $\Omega_n^{\alpha}$ in order to fulfil
\begin{equation}
\label{proportional condition}
\begin{split}
\vec{\beta}_{x,m_x} \propto \vec{v}_c, \\
\vec{\beta}_{y,m_y} \propto \vec{v}_s.
\end{split}
\end{equation}
For concreteness we assume that the two transversal modes and frequencies are equal, i.e. $\vec{b}_{x,n} = \vec{b}_{y,n} =: \vec{b}_{n}$ and $\omega_n^{x} = \omega_n^{y} =: \omega_n$ for $n = 1,\ldots,N$. We choose $q$ in order to obtain condition Eq.\ \eqref{proportional condition} for the two transversal modes that correspond to the second and third highest frequencies, i.e. for $m_x = N-1$ and $m_y = N-2$ (recall that the center of mass mode has the highest frequency for the transversal modes). 
In the example shown in the main text we choose $N=15$ and $z=16$ and find
 \begin{equation}
 \vec{v}_c = \frac{1}{2}\left[\begin{array}{c} \sqrt{2+\sqrt{2}} \\ \sqrt{2} \\ \sqrt{2-\sqrt{2}} \\ 0 \\ -\sqrt{2-\sqrt{2}} \\ -\sqrt{2} \\ -\sqrt{2+\sqrt{2}} \\ -1 \\ -\sqrt{2+\sqrt{2}} \\ -\sqrt{2} \\ -\sqrt{2-\sqrt{2}} \\ 0 \\ \sqrt{2-\sqrt{2}} \\ \sqrt{2} \\ \sqrt{2+\sqrt{2}} \end{array}\right]\text{\; and \;} \vec{v}_s = \frac{1}{2}\left[\begin{array}{c} \sqrt{2-\sqrt{2}} \\ \sqrt{2} \\ \sqrt{2+\sqrt{2}} \\ 1 \\ \sqrt{2+\sqrt{2}} \\ \sqrt{2} \\ \sqrt{2-\sqrt{2}} \\ 0 \\ -\sqrt{2-\sqrt{2}} \\ -\sqrt{2} \\ -\sqrt{2+\sqrt{2}} \\ -1 \\ -\sqrt{2+\sqrt{2}} \\ -\sqrt{2} \\ -\sqrt{2-\sqrt{2}} \end{array}\right]
 \end{equation}
 Two modes that resemble these vectors, i.e.\ that have entries with the same sign, are given by 
 \begin{equation}
 \vec{b}_{13} \propto \left[\begin{array}{c} 1 \\ 0.4756 \\ 0.1256 \\ -0.1245 \\ -0.3028 \\ -0.4231 \\ -0.4929 \\ -0.5157 \\ -0.4929 \\ -4231 \\ -3028 \\ -0.1245 \\ 0.1256 \\ 0.4756 \\ 1 \end{array}\right]\text{\; and \;}  \vec{b}_{14} \propto \left[\begin{array}{c} 1 \\ 0.8091 \\ 0.6509 \\ 0.5085 \\ 0.3752 \\ 0.2475 \\ 0.1230 \\ 0 \\ -0.1230 \\ -0.2475 \\ -0.3752 \\ -0.5085 \\ -0.6509 \\ -0.8091 \\ -1 \end{array}\right].
 \end{equation}
We can thus choose the Rabi frequencies appropriately such that conditition Eq.\ \eqref{proportional condition} is fulfilled, which results in the Hamiltonian
 \begin{equation}
 H \propto \sum_{i,j=1}^{15} \cos\left(\frac{\pi}{8}(i-j)\right) \sigma_z^{i}\sigma_z^{j} + B \sum_{i=1}^{15} \sigma_x^{i}.
 \end{equation}

 \section{Lower bound to the ground state energy}
 
 \label{app:lower bound}
 
 In the following we describe how to obtain a lower bound using semidefinite programming as described in \cite{baumgratz} for the permutation invariant case.
 The ground state energy of a Hamiltonian $H$ may be expressed as the minimization of $\tr[H \varrho]$ over density matrices $\varrho \ge 0$ with $\tr[\varrho] = 1$. Now, consider a set of operators $\{ X_n\}_{n=1}^M$ with the property that 
 \begin{equation}
 \label{Hcoefficients}
 H = \sum_{m,n} \mathcal{H}_{mn} X_m^\dagger X_n
 \end{equation}
  for an $M\times M$ matrix $\mathcal{H}$. Note that, for any density matrix $\varrho$ the matrix $\mathcal{X}$ with entries $\mathcal{X}_{mn} = \tr[ X_{m}^\dagger X_n \varrho]$ is positive semidefinite. Hence, a lower bound to the ground state energy of $H$ is given by $\min_{\mathcal{X}\ge 0} \tr[\mathcal{H} \mathcal{X}]$ where the optimization is over positive semidefinite matrices $\mathcal{X}$ that may follow additional constraints imposed by relations among the operators $\{ X_m^\dagger X_n\}_{m,n}$.

 In the present problem, when we consider spin-1/2 particles, the Hamiltonian is given by

 \begin{equation}
 \label{hamiltoniansecJ}
 H = \frac{1}{4}\sum_{m,n} cos(q(m-n)) \sigma_x^m \sigma_x^n - \frac{1}{2}\sum_{n}(s \cos qn + t \sin qn) \sigma_x^n + s^2+t^2- \frac{B}{2} \sum_n \sigma_z^n.
 \end{equation}

In the following we concentrate on the case $q=0$. The Hamiltonian can be decomposed into a direction sum according the decomposition of the Hilbert space into irreducible representations of $SU(2)$. With $J_x$, $J_y$, and $J_z$ denoting the spin-$J$ operators, we may thus consider
\begin{equation}
H_J = J_x^2 - s J_x + s^2 - \frac{B}{2} J_z.
\end{equation}
Furthermore we define the set $\{X_n\}_n$ to be $\{\id, J_x,J_y,J_z,J_xJ_x,J_xJ_y, \ldots, J_z J_z\}$, i.e. it consist of first and second moments of the the spin operators, and the identity. The operators fulfil the commutator relations $[J_\alpha, J_\beta] = \mi\epsilon_{\alpha\beta\gamma} J_\gamma$, where $\alpha, \beta, \gamma \in \{x,y,z\}$ and $\epsilon_{\alpha\beta\gamma}$ denotes the Levi-Civita symbol, and the relation $J_x^2+J_y^2+J_z^2 = J(J+1) \id$. As mentioned before, we use these relations to linearly constrain the matrix $\mathcal{X}$ defined above, and may denote, generally, the set of all matrices that fulfil these constraints by $\mathcal{B}$. Then the SDP
\begin{equation}
\label{min H_J}
\begin{split}
&\min \; \tr(\mathcal{H}_J\mathcal{X}) \\
& \;\;\;\;\;\;\;\;\; \mathcal{X} \ge 0, \\
& \;\;\;\;\;\;\;\;\; \mathcal{X} \in \mathcal{B},
\end{split}
\end{equation}
where
\begin{equation}
\mathcal{H}_J = \left(\begin{array}{ccccc} 1 & -s & -\frac{B}{2} & \mathbb{O}_{1\times 10} \\ -s & s^2 & 0 & \vdots\\ \frac{B}{2} & 0 & 0 &  \\ \mathbb{O}_{10\times 1} & \cdots &  & \mathbb{O}_{10 \times 10}\\
\end{array}\right),
\end{equation}
provides a lower bound the ground state of $\mathcal{H}_J$ for a fixed value of $s$. For a fixed number of spins, we have to take into account all $J$ in the decomposition of the Hilbert space, i.e.\ $J = \frac{N}{2}, \frac{N}{2} -1, \frac{N}{2} -2, \ldots$ in order to obtain a lower bound for the Hamiltonian Eq.\ \eqref{hamiltoniansecJ}.
Here, $\mathbb{O}_{m\times n}$ denotes the matrix of dimension $m \times n$ with all zero entries. The problem Eq.\ \eqref{min H_J} can be solved using tools from convex optimization \cite{CVX, yalmip}. We use a separate optimization to find the minimum over $s$. Note that the dimension of matrix $\mathcal{X}$ does not increase with $J$. 

We can also use directly the relations among the Pauli matrices without decomposing the Hilbert space into irreducible representations. This will also result in an optimization where the involved matrices are of fixed dimension. This has the advantage, that for a fixed number of spins, we, indeed, only have to take into account a single optimization. To define the set $\{X_n\}_n$ we choose (in the following order) the operators $\sigma_\alpha^l$, $\id$ and $\sigma_\beta^m \sigma_\gamma^n$, with $\alpha = x,y,z$, $(\beta,\gamma) = (x,x),(x,y),(x,z),(y,y),(y,z)$ and $1\le l,m\ne n \le N$. With this choice we may write the coefficient matrix of the Hamiltonian $\mathcal{H}$ as
\begin{equation}
\mathcal{H} = \left(\begin{array}{ccccc} \mathcal{I}_N & -i\frac{B}{2} \id & 0 & 0 &  \\ i\frac{B}{2} \id & 0 & -i\frac{s}{2}\id & 0 & \mathbb{O}_{3N\times N(N-1)} \\ 0 & i\frac{s}{2}\id & 0 & 0 & \\ 0 & 0 & 0 & s^2 & 0 \\
 & \mathbb{O}_{N(N-1)\times 3N} &  & 0 & \mathbb{O}_{N(N-1)\times N(N-1)}
\end{array}\right),
\end{equation}
where $\mathcal{I}_d$ is defined as the matrix of dimension $d\times d$ with all entries equal to 1.
 The moment matrix $\mathcal{X}$ can be parametrized as

\begin{equation}
\label{constraintXblock}
\mathcal{X} = \left(\begin{array}{cccccccccc} \mathcal{X}_{x,x} & \mathcal{X}_{x,y} & \mathcal{X}_{x,z} & \vec{\mathcal{X}}_{x} & \mathcal{X}_{x,xx} & \mathcal{X}_{x,xy} & \mathcal{X}_{x,xz} & \mathcal{X}_{x,yy} & \mathcal{X}_{x,yz} \\ \mathcal{X}_{x,y}^\dag & \mathcal{X}_{y,y} & \mathcal{X}_{y,z} & \vec{\mathcal{X}}_{y} & \mathcal{X}_{y,xx} & \mathcal{X}_{y,xy} & \mathcal{X}_{y,xz} & \mathcal{X}_{y,yy} & \mathcal{X}_{y,yz} \\ \mathcal{X}_{x,z}^\dag & \mathcal{X}_{y,z}^\dag & \mathcal{X}_{z,z} & \vec{\mathcal{X}}_{z}  & \mathcal{X}_{z,xx} & \mathcal{X}_{z,xy} & \mathcal{X}_{z,xz} & \mathcal{X}_{z,yy} & \mathcal{X}_{z,yz} \\ \vec{\mathcal{X}}_{x}^\dag & \vec{\mathcal{X}}_{y}^\dag & \vec{\mathcal{X}}_{z}^\dag & \mathcal{X}_{1,1} & \vec{\mathcal{X}}_{xx}^\dagger & \vec{\mathcal{X}}_{xy}^\dagger & \vec{\mathcal{X}}_{xz}^\dagger & \vec{\mathcal{X}}_{yy}^\dagger & \vec{\mathcal{X}}_{yz}^\dagger \\ \mathcal{X}_{x,xx}^\dag & \mathcal{X}_{y,xx}^\dag & \mathcal{X}_{z,xx}^\dag & \vec{\mathcal{X}}_{xx} & \mathcal{X}_{xx,xx} & \mathcal{X}_{xx,xy} & \mathcal{X}_{xx,xz} & \mathcal{X}_{xx,yy} & \mathcal{X}_{xx,yz} \\ \mathcal{X}_{x,xy}^\dag &\mathcal{X}_{y,xy}^\dag &\mathcal{X}_{z,xy}^\dag & \vec{\mathcal{X}}_{xy} & \mathcal{X}_{xx,xy}^\dag & \mathcal{X}_{xy,xy} & \mathcal{X}_{xy,xz} & \mathcal{X}_{xy,yy} & \mathcal{X}_{xy,yz} \\ \mathcal{X}_{x,xz}^\dag &\mathcal{X}_{y,xz}^\dag &\mathcal{X}_{z,xz}^\dag & \vec{\mathcal{X}}_{xz} & \mathcal{X}_{xx,xz}^\dag & \mathcal{X}_{xy,xz}^\dag & \mathcal{X}_{xz,xz} & \mathcal{X}_{xz,yy} & \mathcal{X}_{xz,yz} \\ \mathcal{X}_{x,yy}^\dag &\mathcal{X}_{y,yy}^\dag &\mathcal{X}_{z,yy}^\dag & \vec{\mathcal{X}}_{yy} & \mathcal{X}_{xx,yy}^\dag & \mathcal{X}_{xy,yy}^\dag & \mathcal{X}_{xz,yy}^\dag & \mathcal{X}_{yy,yy} & \mathcal{X}_{yy,yz} \\ \mathcal{X}_{x,yz}^\dag &\mathcal{X}_{y,yz}^\dag &\mathcal{X}_{z,yz}^\dag & \vec{\mathcal{X}}_{yz} & \mathcal{X}_{xx,yz}^\dag & \mathcal{X}_{xy,yz}^\dag & \mathcal{X}_{xz,yz}^\dag & \mathcal{X}_{yy,yz}^\dag & \mathcal{X}_{yz,yz} \end{array} \right),
\end{equation}

where the block matrices are of the form $\mathcal{X}_{\alpha,\beta} \in \cc^{N\times N}$, $\vec{\mathcal{X}}_{\alpha} \in \cc^{N}$, $\mathcal{X}_{\alpha,\beta\gamma} \in \cc^{N\times N(N-1)}$ and $\mathcal{X}_{\alpha\beta,\gamma\delta} \in \cc^{N(N-1) \times N(N-1)}$ and every block represents a moment matrix with entries $\mathcal{X}_{\alpha,\beta} = \langle \hat\sigma_{\alpha}\hat\sigma_{\beta} \rangle$, $\mathcal{X}_{\alpha,\beta\gamma} = \langle \hat\sigma_{\alpha}\hat\sigma_{\beta}\hat\sigma_{\gamma} \rangle$ and $\mathcal{X}_{\alpha\beta,\gamma\delta} = \langle \hat\sigma_{\alpha}\hat\sigma_{\beta}\hat\sigma_{\gamma}\hat\sigma_{\delta} \rangle$, respectively. The blocks are linked with each other through the algebraic relations $\sigma_\alpha \sigma_\beta = \delta_{\alpha,\beta} + \sum_{\gamma={x,y,z}} \mi \epsilon_{\alpha\beta\gamma} \sigma_\gamma$ that we translate into constraints on $\mathcal{X}$. For example, $(\mathcal{X}_{x,xx})_{l,mn} = \sigma_x^l\sigma_x^m\sigma_x^n$ and, also, $(\mathcal{X}_{xy,xz})_{l,m,n,m} = \mi \sigma_x^l\sigma_x^m\sigma_x^n$, for $l\ne n$ and $l \ne m$. Therefore, we require that $\sum_{l\ne n,l \ne m} (\mathcal{X}_{x,xx})_{l,mn} + \mi (\mathcal{X}_{xy,xz})_{l,m,n,m} = 0$. We define the matrix $\mathcal{J}_d := \mathcal{I}_d - \id_{d\times d}$. Furthermore, let $\mathcal{A}^{(n)}$ and $\mathcal{B}^{(n)}$, $n = 1,\ldots, 6,$ be the matrices with entries, respectively, given by

\begin{equation}
(\mathcal{A}^{(1)})_{l,mn} = \delta_{l,m}, \;\;\;\;\;\;\; (\mathcal{A}^{(2)})_{l,mn} = \delta_{l,n}, \;\;\;\;\;\;\; (\mathcal{A}^{(3)})_{l,mn} = 1-\delta_{l,m}-\delta_{l,n}, \end{equation}
and
\begin{equation}
\begin{split}
(\mathcal{B}^{(1)})_{kl,mn} = \delta_{k,m}\delta_{l,n}, \;\;\;\;\;\;\; (\mathcal{B}^{(2)})_{kl,mn} = \delta_{k,m}(1-\delta_{l,n}), \;\;\;\;\;\;\; (\mathcal{B}^{(3)})_{kl,mn} = (1-\delta_{k,m})\delta_{l,n}, \\
(\mathcal{B}^{(4)})_{kl,mn} = \delta_{k,n}\delta_{l,m}, \;\;\;\;\;\;\; (\mathcal{B}^{(5)})_{kl,mn} = \delta_{k,n}(1-\delta_{l,m}), \;\;\;\;\;\;\; (\mathcal{B}^{(6)})_{kl,mn} = (1-\delta_{k,n})\delta_{l,m},
\end{split}
\end{equation}

and $\mathcal{B}^{(7)} = \mathcal{I}_{N(N-1)} - \sum_{n=1}^6 \mathcal{B}^{(n)}$. Next, we introduce the matrix $\mathcal{W}$ with the same block structure as Eq.\ \eqref{constraintXblock} and blocks of the form $\mathcal{W}_{\alpha,\beta} = w_{\alpha,\beta}^{(1)} \id + w_{\alpha,\beta}^{(2)} \mathcal{J},$ $\mathcal{W}_{\alpha,\beta\gamma} = \sum_{n=1}^6 w_{\alpha,\beta\gamma}^{(n)} \mathcal{A}^{(n)}$ and $\mathcal{W}_{\alpha\beta,\gamma\delta} = \sum_{n=1}^6 w_{\alpha\beta,\gamma\delta}^{(n)} \mathcal{B}^{(n)}$, respectively, and we require that

\begin{equation}
\label{conditions}
\begin{split}
0 &= w_{yz,xz}^{(2)} = w_{xz,xz}^{(7)} = w_{yz,yz}^{(7)} = w_{yz,xy}^{(7)} = w_{xy,xy}^{(7)} = w_{xx,xx}^{(7)} = w_{yy,yy}^{(7)}, \\
0 &= w_{x,xx}^{(3)} + \mi w_{xy,xz}^{(3)}, \\
0 &= w_{yz,xy}^{(7)} + w_{yy,xz}^{(7)}, \\
0 &= w_{xy,xy}^{(7)} + 2 w_{xx,yy}^{(7)}, \\
0 &= -\mi w_{yz,xx}^{(2)} - \mi w_{yz,xx}^{(5)} + w_{z,xz}^{(3)} - \mi w_{xy,xz}^{(6)}, \\
0 &= w_{y,yz}^{(3)} + w_{z,yy}^{(3)} + \mi w_{xy,yy}^{(2)} + \mi w_{xy,yy}^{(5)} + \mi w_{yz,xz}^{(6)}, \\
0 &= w_{z,xx}^{(3)} - \mi w_{xy,xx}^{(3)} - \mi w_{xy,xx}^{(6)} + w_{x,xz}^{(3)} + \mi w_{yz,xz}^{(5)}, \\
0 &= \mi w_{xyy}^{(3)} + w_{y,xy}^{(3)} + \mi w_{yz,xx}^{(3)} + \mi w_{yz,xx}^{(6)} - \mi w_{yz,yy}^{(3)} - \mi w_{yz,yy}^{(6)} - \mi w_{xy,xz}^{(5)}, \\
0 &= w_{x,z}^{(2)} - \mi w_{z,yz}^{(1)} + \mi w_{x,xy}^{(2)} + w_{yz,xy}^{(1)} - \mi w_{y,xx}^{(1)} - \mi w_{y,xx}^{(2)} + (n-2) w_{yz,xy}^{(5)}, \\
0 &= w_x + \mi w_{y,z}^{(1)} + (N-1)\left(w_{y,xy}^{(2)} + w_{x,xx}^{(1)} + w_{x,xx}^{(2)} - w_{yz,yy}^{(1)} - \mi w_{yz,yy}^{(4)} + w_{z,xz}^{(2)} + \mi w_{xyxz}^{(1)}\right), \\
0 &= w_z + \mi w_{x,y}^{(1)} + (N-1)\left(w_{y,yz}^{(1)} - \mi w_{xy,xx}^{(1)} - \mi w_{xy,xx}^{(4)} + \mi w_{xy,yy}^{(1)} + \mi w_{xy,yy}^{(4)} + w_{x,xz}^{(1)} - \mi w_{yz,xz}^{(1)}\right), \\
0 &= w_{z,z}^{(2)} + w_{xy,xy}^{(4)} + 2 \mi \left(w_{x,yz}^{(1)} - w_{xx,yy}^{(1)} - w_{xx,yy}^{(1)} - w_{xx,yy}^{(4)} - \mi w_{y,xz}^{(1)}\right) + (N-2) \left(w_{yz,yz}^{(2)} + w_{xz,xz}^{(2)}\right), \\
0 &= w_{y,y}^{(2)} + w_{xz,xz}^{(4)} - 2\mi \left(w_{x,yz}^{(2)} + \mi w_{z,xy}^{(1)} + w_{yy}\right) + (N-2)\left(w_{yz,yz}^{(3)} +  w_{xy,xy}^{(2)} + w_{yy,yy}^{(2)} + w_{yy,yy}^{(3)} + w_{yy,yy}^{(5)} + w_{yy,yy}^{(6)}\right), \\
0 &= w_{x,x}^{(2)} + w_{yz,yz}^{(4)}- 2\left(w_{xx} + \mi w_{z,xy}^{(2)} + \mi w_{y,xz}^{(2)}\right) + (N-2)\left(w_{xy,xy}^{(3)} + w_{xx,xx}^{(2)} + w_{xx,xx}^{(3)} + w_{xx,xx}^{(5)} + w_{xx,xx}^{(6)} + w_{xz,xz}^{(3)}\right), \\
0 &= w_{x,z}^{(2)} - \mi w_{z,yz}^{(1)} + \mi w_{x,xy}^{(2)} + w_{yz,xy}^{(1)} - \mi w_{y,xx}^{(1)} - \mi w_{y,xx}^{(2)} + w_{xz} + w_{xx,xz}^{(2)} + w_{xx,xz}^{(6)} + w_{yy,xz}^{(1)} + w_{yy,xz}^{(4)} + (n-2)w_{yz,xy}^{(5)}. \\
\end{split}
\end{equation}

By the Pauli algebraic relations, the conditions Eq.\ \eqref{conditions} guarantee that $\tr[\mathcal{W}\mathcal{X}] = N \sum w_{\alpha,\alpha}^{(1)} + 1 + N(N-1)\sum w_{\alpha\beta,\alpha\beta}^{(1)}$, where the last sum runs over pairs $(\alpha,\beta) = (x,x),(x,y),(x,z),(y,y),(y,z)$. Therefore, for all $\mathcal{X}$

\begin{equation}
\tr[\mathcal{H}\mathcal{X}] \ge \max \left\{ N \sum w_{\alpha,\alpha}^{(1)} + 1 + N(N-1)\sum w_{\alpha\beta,\alpha\beta}^{(1)}:\mathcal{H} \ge \mathcal{W}\right\},
\end{equation}

where the maximum is over $w_{\alpha,\beta}^{(n)},$ $w_{\alpha,\beta\gamma}^{(n)}$ and $w_{\alpha\beta,\gamma\delta}^{(n)}$.

Now, we set out to show that the eigenvalues of $\mathcal{H} - \mathcal{W}$ can be determined efficiently, i.e.\ the problem reduces to determine the eigenvalues of a $32 \times 32$ matrix. We start by explicitly constructing a basis. It will be useful that we can represent any vector $\vec{v} \in \cc^{N(N-1)}$ with entries $v_n$, $n=1,\ldots,N(N-1)$ equivalently by the matrix
\begin{equation}
\label{reshaping}
\underline{v} = \left(\begin{array}{cccc} 0 & v_N & \cdots & v_{(N-1)^2+1} \\
v_1 & 0 & & v_{(N-1)^2+2} \\
\vdots & \vdots & \ddots  & \vdots \\  v_{N-1} & v_{2(N-1)} & \cdots & 0 \end{array} \right). 
\end{equation}

 In the following $\left\{\vec{e}_j^d\right\}_{j=1}^d$ will denote the standard basis of $\cc^d$. With this, we introduce $\mathcal{E}^d_{j,k} = \vec{e}^d_j (\vec{e}_k^{d})^T$. Now for any pair $(k,l)$ we define
\begin{equation}
\label{permuterowseven}
\mathcal{T}_{j,k} = \left(\begin{array}{cc} \id_{d} - \mathcal{E}_{j,j}^{d} & \mathcal{E}_{j,k}^{d} \\ \mathcal{E}_{k,j}^{d} & \id_{d} - \mathcal{E}_{k,k}^{d} \end{array}\right)
\end{equation}
and 
\begin{equation}
\label{permuterowsodd}
\mathcal{T}_{j,k} = \left(\begin{array}{ccc} \id_{d} - \mathcal{E}_{j,j}^{d} & 0 & \mathcal{E}_{j,k}^{d} \\ 0 & 1 & 0 \\ \mathcal{E}_{k,j}^{d} & 0 & \id_{d} - \mathcal{E}_{k,k}^{d} \end{array}\right),
\end{equation}
for $N$ even and odd, respectively. Now, let $v_{j,k}^{(n)}$, $n = 1,2,3$ be the vectors with matrix representation given by ($d = \lfloor N/2\rfloor$)

\begin{equation}
\label{vectors123even}
\underline{v}^{(1)} = \frac{1}{\sqrt{\mathcal{N}_1}}\mathcal{J}_N, \;\;\;\;\; \underline{v}_{j,k}^{(2)} = \frac{1}{\sqrt{\mathcal{N}_2}} \mathcal{T}_{j,k}\left(\begin{array}{cc} \mathcal{J}_{d} & 0 \\ 0 & -\mathcal{J}_{d} \end{array}\right) \mathcal{T}_{j,k}, \;\;\;\;\; \underline{v}_{j,k}^{(3)} = \frac{1}{\sqrt{\mathcal{N}_3}}\mathcal{T}_{j,k}\left(\begin{array}{cc} 0 & \mathcal{I}_{d} \\ -\mathcal{I}_{d} & 0 \end{array}\right)\mathcal{T}_{j,k},
\end{equation}
if $N$ is even, where $\mathcal{N}_1=N(N-1)$, $\mathcal{N}_2=N(N/2-1)$ and $\mathcal{N}_3=N^2/2$, and 
\begin{equation}
\label{vectors123odd}
\underline{v}^{(1)} = \frac{1}{\sqrt{\mathcal{N}_1}}\mathcal{J}_N, \;\;\;\;\; \underline{v}_{j,k}^{(2)} = \frac{1}{\sqrt{\mathcal{N}_2}} \mathcal{T}_{j,k} \left(\begin{array}{ccc} \mathcal{J}_{d} & \frac{1}{2}\vec{1}_{d} & 0 \\ \frac{1}{2}\vec{1}_{d}^T & 0 & -\frac{1}{2}\vec{1}_{d}^T \\ 0 & -\frac{1}{2}\vec{1}_{d} & -\mathcal{J}_{d} \end{array}\right) \mathcal{T}_{j,k}, \;\;\;\;\; \underline{v}_{j,k}^{(3)} = \frac{1}{\sqrt{\mathcal{N}_3}}\mathcal{T}_{j,k} \left(\begin{array}{ccc} 0 & \frac{1}{2}\vec{1}_{d} & \mathcal{I}_{d}\\ -\frac{1}{2}\vec{1}_{d}^T & 0 & \frac{1}{2}\vec{1}_{d}^T \\ -\mathcal{I}_{d} & -\frac{1}{2}\vec{1}_{d} & 0 \end{array}\right)\mathcal{T}_{j,k}
\end{equation}
if $N$ is odd, where $\mathcal{N}_1=N(N-1)$, $\mathcal{N}_2=(N-1)(N-2)/2$ and $\mathcal{N}_3=N(N-1)/2$. For $N$ odd we additionally define 
\begin{equation}
\label{v1Nhalbe}
\underline{v}_{1,\lfloor\frac{N}{2}\rfloor}^{(2)} = \frac{1}{\sqrt{\mathcal{N}_2}} \mathcal{T}_{1,\lfloor\frac{N}{2}\rfloor} \left(\begin{array}{ccc} \mathcal{J}_{d} & \frac{1}{2}\vec{1}_{d} & 0 \\ \frac{1}{2}\vec{1}_{d}^T & 0 & -\frac{1}{2}\vec{1}_{d}^T \\ 0 & -\frac{1}{2}\vec{1}_{d} & -\mathcal{J}_{d} \end{array}\right) \mathcal{T}_{1,\lfloor\frac{N}{2}\rfloor}, \;\;\;\;\; \underline{v}_{1,\lfloor\frac{N}{2}\rfloor}^{(2)} = \frac{1}{\sqrt{\mathcal{N}_3}}\mathcal{T}_{0,N/2} \left(\begin{array}{ccc} 0 & \frac{1}{2}\vec{1}_{d} & \mathcal{I}_{d}\\ -\frac{1}{2}\vec{1}_{d}^T & 0 & \frac{1}{2}\vec{1}_{d}^T \\ -\mathcal{I}_{d} & -\frac{1}{2}\vec{1}_{d} & 0 \end{array}\right)\mathcal{T}_{1,\lfloor\frac{N}{2}\rfloor},
\end{equation}

where
\begin{equation}
\label{T1Nhalbe}
\mathcal{T}_{1,\lfloor\frac{N}{2}\rfloor} = \left(\begin{array}{ccc} \id_{d} - \mathcal{E}_{1,1}^{d} & \vec{e}_1^{d} & 0 \\ \vec{e}_1^{dT} & 0 & 0 \\ 0 & 0 & \id_{d} \end{array}\right).
\end{equation}
Furthermore, let

\begin{equation}
\label{cossinbasis}
\begin{split}
\underline{w}_{j,k;1}^{(4)} &= \vec{c}_k\vec{c}_l^T + \vec{c}_l\vec{c}_k^T + \vec{s}_k\vec{s}_l^T + \vec{s}_l\vec{s}_k^T,\;\;\;\;\;\;k\le l,\;k,l = 1,\ldots, \left\lfloor\frac{N}{2}\right\rfloor, \\
\underline{w}_{j,k;2}^{(4)} &= \vec{c}_k\vec{c}_l^T + \vec{c}_l\vec{c}_k^T - \vec{s}_k\vec{s}_l^T -  \vec{s}_l\vec{s}_k^T,\;\;\;\;\;\;k\le l,\;k,l = 1,\ldots, \left\lfloor\frac{N-1}{2}\right\rfloor, \\
\underline{w}_{j,k;3}^{(4)} &= \vec{c}_k\vec{s}_l^T + \vec{s}_k\vec{c}_l^T + \vec{c}_l\vec{s}_k^T + \vec{s}_l\vec{c}_k^T,\;\;\;\;\;\;k \le l,\; k,l = 1,\ldots, \left\lfloor\frac{N-1}{2}\right\rfloor, \\
\underline{w}_{j,k;4}^{(4)} &= \vec{c}_k\vec{s}_l^T + \vec{s}_l\vec{c}_k^T - \vec{c}_l\vec{s}_k^T - \vec{s}_k\vec{c}_l^T,\;\;\;\;\;\;k < l,\;k,l = 1,\ldots, \left\lfloor\frac{N}{2}\right\rfloor,
\end{split}
\end{equation}
and
\begin{equation}
\label{cossinantisymmetric}
\begin{split}
\underline{w}_{j,k;1}^{(5)} &= \vec{c}_k \vec{c}_l^T - \vec{c}_l \vec{c}_k^T, \\
\underline{w}_{j,k;2}^{(5)} &= \vec{c}_k \vec{s}_l^T - \vec{s}_l \vec{c}_k^T, \\
\underline{w}_{j,k;3}^{(5)} &= \vec{s}_k \vec{s}_l^T - \vec{s}_l \vec{s}_k^T. \\
\end{split}
\end{equation}
By Lemma \ref{lemmabasis} below, we can form linear combinations of, respectively, $\underline{v}_{j,k}^{(n)}$ for $n=1,2,3$ and $\underline{w}_{j,k;n}^{(n)}$ for $n=4,5$ so that the corresponding vectors form an orthonormal basis $\{\tilde{\vec{v}}_j^{(n)}\}_{j,n}$ of $\cc^{N(N-1)}$. Additionally, $\vec{u}_j := (\mathcal{A}^{(1)}+\mathcal{A}^{(2)}) \tilde{\vec{v}}_j^{(2)}/\|(\mathcal{A}^{(1)}+\mathcal{A}^{(2)}) \tilde{\vec{v}}_j^{(2)}\|$ defines a basis of $\cc^N$. We find
\begin{equation}
\label{block diagonal form}
\begin{split}
u_j^T \mathcal{J} u_j = (N-1)\delta_{1,j}, \;\;\; u_j^T\mathcal{A}^{(1)} \tilde{\vec{v}}_j^{(n)} = \sqrt{2(N-2)} \delta_{n,2}\delta_{j,k}, \;\;\; u_j^T\mathcal{A}^{(2)} \tilde{\vec{v}}_j^{(n)} = -\sqrt{2(N-2)} \delta_{n,2}\delta_{j,k} \\
u_j^T \vec{1} = \sqrt{N}\delta_{j,1}, \;\;\; \tilde{\vec{v}}_j^{(n)T}\vec{1} = \sqrt{N-1}\delta_{n,1} \hspace{4cm} \\
\tilde{\vec{v}}_j^{(n)} \mathcal{B}^{(2)} \tilde{\vec{v}}_k^{(m)} = (N-1)\delta_{j,k}\delta_{n,1} \;\;\;\;\; m,n=1,4,5 \hspace{3.4cm} \\
(\tilde{\vec{v}}_j^{(n)} \mathcal{B}^{(2)} \tilde{\vec{v}}_k^{(m)})_{m=2,3;n=2,3} = \delta_{j,k} \left( \begin{array}{cc} \alpha & -\beta \\ -\beta & \gamma \end{array} \right), \;\;\; (\tilde{\vec{v}}_j^{(n)} \mathcal{B}^{(3)} \tilde{\vec{v}}_k^{(m)})_{m=2,3;n=2,3} = \delta_{j,k} \left( \begin{array}{cc}  \alpha & \beta \\ \beta & \gamma \end{array} \right), \hspace{0.2cm} \\
(\tilde{\vec{v}}_j^{(n)} \mathcal{B}^{(5)} \tilde{\vec{v}}_k^{(m)})_{m=2,3;n=2,3} = \delta_{j,k} \left( \begin{array}{cc} -\alpha & -\beta \\ \beta & \gamma \end{array} \right), \;\;\; (\tilde{\vec{v}}_j^{(n)} \mathcal{B}^{(6)} \tilde{\vec{v}}_k^{(m)})_{m=2,3;n=2,3} = \delta_{j,k} \left( \begin{array}{cc} -\alpha & \beta \\ -\beta & \gamma  \end{array} \right),
\end{split}
\end{equation}

Since all matrix elements in Eq.\ \eqref{block diagonal form} are proportional to $\delta_{j,k}$ we can form a basis of $\cc^{3N+1+5N(N-1)}$ with $\mathcal{H} - \mathcal{W}$ block diagonal and with all blocks equal. Each block is of size $32 \times 32$.

We split the proof of Lemma \ref{lemmabasis} by considering first two preparatory Lemmas.

\begin{lemma}
	(i) There are linear combinations $\underline{\tilde{v}}_l^{(4)} = \sum_{j,k,m} c^l_{jkm} \underline{w}_{j,k;m}^{(4)}$ for $l=1,\ldots,N-1,$ such that the vectors $\tilde{\vec{v}}_j^{(4)}$ defined via Eq.\ \eqref{reshaping} are orthonormal and span a $(N-1)(N-2)/2-1$-dimensional subspace. Furthermore, (ii) the vectors corresponding to $\underline{w}_{j,k;m}^{(5)}$ span a $(N-1)(N-2)/2$-dimensional subspace.
\end{lemma}

\proof{ (i) The diagonal entries of $\underline{w}_{j,k;m}^{(4)}$ for $m=1,\ldots,4$ are given by
	\begin{equation}
	\begin{split}
	\mathrm{diag}(\underline{w}_{j,k;1}^{(4)})_l &= (\vec{c}_{j-k})_l, \\
	\mathrm{diag}(\underline{w}_{j,k;2}^{(4)})_l &= (\vec{c}_{j+k})_l, \\
	\mathrm{diag}(\underline{w}_{j,k;3}^{(4)})_l &= (\vec{s}_{j-k})_l, \\
	\mathrm{diag}(\underline{w}_{j,k;4}^{(4)})_l &= (\vec{s}_{j+k})_l.
	\end{split}
	\end{equation}
	Since $\{\{\vec{c}_j\}_j,\{\vec{s}_j\}_j\}$ are mutually orthogonal, we may consider linear combinations of $\underline{w}_{j,k;m}^{(4)}$ that have equal diagonals. Counting reveals that these matrices may have one of $N$ distinct diagonals. It thus follows by the fact that $\underline{w}_{j,k;m}^{(4)}$ are mutually orthogonal in the Hilbert-Schmidt inner product, that there are $(N-1)(N-2)/2-1$ linear independent linear combinations with zeros on the diagonal. (ii) The matrices $\underline{w}_{j,k;m}^{(5)}$ are mutually orthogonal with zeros on the diagonal. Hence, counting shows that they span a subspace of dimension $(N-1)(N-2)/2$.
	}

\begin{lemma}
	There are coefficients $c^l_{jk}$ such that $\underline{\tilde{v}}_l^{(n)} = \sum_{j,k} c_{jk}^l \underline{{v}}_{jk}^{(n)}$ define two sets of orthonormal vectors $\{\tilde{\vec{v}}_j^{(n)}\}_j$, for $n=2,3$, via the correspondence in Eq.\ \eqref{reshaping}, that, respectively, span $(N-1)$-dimensional subspaces.
\end{lemma}

\proof{In order to show that $\vec{v}_{j,k}^{(2)}$ and $\vec{v}_{j,k}^{(3)}$ defined through the Eqs.\ \eqref{vectors123even}, \eqref{vectors123odd} and \eqref{v1Nhalbe} span a $(N-1)$-dimensional subspace we calculate the Gramian matrices of the two sets of vectors. Defining for any $N$
	\begin{equation}
	\label{gram23geven}
	 G = \frac{1}{d}\left((d-4) \mathcal{I}_{d}\otimes \mathcal{I}_{d} + 2 (\id_{d} \otimes \mathcal{I}_{d} + \mathcal{I}_{d} \otimes \id_{d^2})\right),
	 \end{equation}
	 with $d = \lfloor N/2 \rfloor$, we find for $N$ even that $(G)_{jk,lm} = \langle \vec{v}_{j,k}^{(2)}, \vec{v}_{l,m}^{(2)} \rangle = \langle \vec{v}_{j,k}^{(3)}, \vec{v}_{l,m}^{(3)} \rangle $. 
	An explicit calculation of the eigendecomposition of $G$ reveals that it is of rank $2d-1 = N-1$.
	For $N$ odd we find that the Gramian matrices for both sets of vectors again coincide and are given by (in the following the first row and column correspond to $\underline{v}_{1,d}^{(2)((3))}$)
	\begin{equation}
	\mathcal{G} = \left( \begin{array}{cc} 1 & \vec{g}^{T} \\ \vec{g} & {G}  \end{array} \right),
	\end{equation}
	where
	\begin{equation}
	\begin{split}
	\vec{g} &= \frac{2}{(N-1)}\left(\frac{N-6}{2} \vec{1}_{d}\otimes\vec{1}_{d}+ \vec{e}_1^{d}\otimes \vec{1}_{d}\right) \\
	\end{split}
	\end{equation}
	and $G$ given in Eq.\ \eqref{gram23geven}. Since $G$ has rank $N-2$, it follows that $\mathcal{G}$ is of rank $N-1$.
	The two sets, for $n=2$ and $n=3$, have identical Gramian matrices and hence the statement follows.}

\begin{lemma}
	\label{lemmabasis}
	The set $\{\tilde{v}_j^{(n)}\}_{j,n}$ is an orthonormal basis of $\cc^{N(N-1)}$.
\end{lemma}

\proof{ For $n=1,\ldots,5$, $\tilde{\vec{v}}_j^{(n)}$ are eigenvectors of $\mathcal{B}^{(2)}+\mathcal{B}^{(3)}$ to the eigenvalue $\lambda_n$, where $\lambda_1 = 2(N-2), \lambda_2 = N-4, \lambda_3 = N-2$ and $\lambda_{4,5} = -2$. In addition, these vectors are eigenvectors of $\mathcal{B}^{(4)}$ to the eigenvalues $\tau_n$, with $\tau_{1,2,4} = 1$ and $\tau_{3,5} = -1$. Thus $\tilde{\vec{v}}_j^{(n)}$ mutually orthogonal for all $j,n$.
	}


\begin{thebibliography}{99}
	\bibitem{Wineland} D.\ Wineland, et al., Spin squeezing and reduced quantum noise in spectroscopy. Phys.\ Rev.\ A 46, R6797 (1992).
	\bibitem{Kitagawa} M.\ Kitagawa, and M.\ Ueda, Squeezed Spin States, Phys.\ Rev.\ A 47, 5138 (1993).
	\bibitem{Ma} J.\ Ma, X.\ Wang, C.P.\ Sun, and F.\ Nori, Quantum Spin Squeezing, Phys.\ Rep.\ 509, 89 (2011).
	\bibitem{Korbicz} J.\ Korbicz, J.\ Cirac, M.\ Lewenstein, Spin Squeezing Inequalities and Entanglement of $N$ Qubit States, Phys.\ Rev.\ Lett.\ 95, 120502 (2005).
	\bibitem{Toth2009} G.\ T\'oth, C.\ Knapp, O.\ G\"uhne, H.J.\ Briegel, Spin Squeezing and Entanglement, Phys.\ Rev.\ A 79, 042334 (2009).
	\bibitem{sorensen} A.S.\ S\o rensen, and K. M\o lmer, Entanglement and Extreme Spin Squeezing, Phys.\ Rev.\ Lett.\ 86, 4431 (2001).
	\bibitem{Vitagliano}
	G. Vitagliano, I. Apellaniz, I.L. Egusquiza, and G. T\'oth, Spin Squeezing and Entanglement for an Arbitrary Spin, Phys.\ Rev.\ A 89, 032307 (2014).
	\bibitem{Vitagliano3}
	G.\ Vitagliano, et al., Spin Squeezing Inequalities for Arbitrary Spin, Phys.\ Rev.\ Lett.\ 107, 240502 (2011).
	\bibitem{blatt}
	R.\ Blatt, and C.F.\ Roos, Quantum simulations with trapped ions, Nat.\ Phys.\ 8, 277 (2012).
	\bibitem{bloch}
	I.\ Bloch, J.\ Dalibard, and S.\ Nascimb\'ene, Quantum simulations with ultracold quantum gases, Nat.\ Phys.\ 8, 267 (2012).
	\bibitem{devoret}
	M.H.\ Devoret, and R.J.\ Schoelkopf, Superconducting Circuits for Quantum Information: An Outlook, Science 339, 1169 (2013).
	\bibitem{cramer}
	M.\ Cramer, H.\ Wunderlich, and M.B.\ Plenio, Measuring Entanglement in Condensed Matter Systems, Phys. Rev. Lett. 106, 020401 (2011).
	\bibitem{marty}
	O. Marty, et al., Quantifying entanglement with scattering experiments, Phys. Rev. B 89, 125117 (2014).
	\bibitem{luecke} B.\ Lücke, J.\ Peise, G.\ Vitagliano, J.\ Arlt, L.\ Santos, G. T\'oth, and C. Klempt, Detecting Multiparticle Entanglement of Dicke States, Phys.\ Rev.\ Lett.\ 112, 155304 (2014).
	\bibitem{Vitagliano2}
	G.\ Vitagliano, et al., Entanglement and extreme spin squeezing of unpolarized states, New J.\ Phys.\ 19, 013027 (2017).
	\bibitem{krammer}
	P.\ Krammer, H.\ Kampermann, D.\ Bruß, R.A.\ Bertlmann, L.C. Kwek, and C.\ Macchiavello, Multipartite Entanglement Detection via Structure Factors, Phys.\ Rev.\ Lett.\ 103, 100502 (2009).
	\bibitem{baumgratz} T.\ Baumgratz, and M.B.\ Plenio, Lower bounds for ground states of condensed matter systems, New J.\ Phys.\ 14, (2012).
	\bibitem{damm} L.\ Dammeier, et al., Uncertainty relations for angular momentum, New J. Phys. 17, 093046 (2015); see also Apellaniz et. al, Optimal witnessing of the quantum Fisher information with few measurements, Phys. Rev. A 95, 032330 (2017).
	\bibitem{mengi} E.\ Mengi, et al., Numerical Optimization of Eigenvalues of Hermitian Matrix Functions, SIAM 35, 2 (2014).
	\bibitem{guehne} O.\ G\"uhne, G.\ T\'oth, and H.J.\ Briegel, Multipartite entanglement in spin chains, New J.\ Phys.\ 7, 229 (2005).
	\bibitem{Woelk} S.\ W\"olk, and O.\ G\"uhne, Characterizing the Width of Entanglement, New J.\ Phys.\ 18, 123024 (2016).
	\bibitem{hart} R.A.\ Hart et al., Observation of antiferromagnetic correlations in the Hubbard model with ultracold atoms, Nature 519, 211 (2015).
	\bibitem{mazurenko} A.\ Mazurenko, et al., Experimental realization of a long-range antiferromagnet in the Hubbard model with ultracold atoms, Nature 545, 462 (2017).
	\bibitem{riedel} M.A.\ Riedel, et al., Atom-chip-based generation of entanglement for quantum metrology, Nature 464, 1170 (2010).
	\bibitem{gross} C.\ Gross, et al., Nonlinear atom interferometer surpasses classical precision limit, Nature 464, 1165 (2010).
	\bibitem{monz} T.\ Monz, et al., 14-Qubit Entanglement: Creation and Coherence, Phys.\ Rev.\ Lett.\ 106, 130506 (2011).
	\bibitem{barreiro} J.T.\ Barreiro, et al., Demonstration of genuine multipartite entanglement with device-independent witnesses, Nat.\ Phys.\ 9, 559 (2013).
	\bibitem{haas} F.\ Haas, et al., Entangled states of more than 40
	atoms in an optical fiber cavity, Science 344, 180 (2014). 
	\bibitem{mcconnell} R.\ McConnell, et al., Entanglement with negative Wigner function of almost 3,000 atoms heralded by one photon, Nature 519, 439 (2015).
	\bibitem{hosten} O.\ Hosten, Measurement noise 100 times lower than the
	quantum-projection limit using entangled atoms, Nature 529, 505 (2016).
	\bibitem{cox} K.C.\ Cox, Deterministic Squeezed States with Collective Measurements and Feedback, Phys.\ Rev.\ Lett.\ 116, 093602 (2016).
	\bibitem{engels} N.J.\ Engels, et al., Bell Correlations in Spin-Squeezed States of 500000 Atoms, Phys.\ Rev.\ Lett.\ 118, 140401 (2017).
	\bibitem{dellantonio} L.\ Dellantonio, et al., Multi-partite entanglement detection with non symmetric probing, Phys.\ Rev.\ A 95, 040301(R) (2017).
	\bibitem{Jensen} J.\ Jensen, and A.R.\ Mackintosh, Rare Earth Magnetism: Structures and Exciations, Clarendon Press, Oxford (1991).
	\bibitem{vitagliano3} 	G.\ Vitagliano, et al., Entanglement and extreme planar spin squeezing, arXiv:1705.09090 (2017).
	\bibitem{sawada}
	J.\ Sawada, Generating bracelets in constant amortized time, SIAM J. Comput., 31(1), 259–268 (2001); S.\ Karima, J.\ Sawada, Z.\ Alamgir, S.M.\ Husnine, Generating bracelets with fixed content, Theoret. Comp. Sc. 475 103, (2013).
	\bibitem{Hauke2015}	P.\ Hauke, L.\ Bonnes, M.\ Heyl and W.\ Lechner, Probing Entanglement in Adiabatic Quantum Optimization with Trapped Ions, Front.\ Phys.\ 3:21, 1 (2015).
	\bibitem{hauke} P.\ Hauke, D.\ Marcos, M.\ Dalmonte, and P.\ Zoller, Quantum Simulation of a Lattice Schwinger Model in a Chain of Trapped Ions, Phys.\ Rev.\ X 3, 041018 (2013).
	\bibitem{lin} G.-D. Lin,\ C.\ Monroe, and L.-M.\ Duan, Sharp Phase Transitions in a Small Frustrated Network of Trapped Ion Spins, Phys.\ Rev.\ Lett.\ 106, 230402 (2011).
	\bibitem{chiuri} A.\ Chiuri, G.\ Vallone, N.\ Bruno, C.\ Macchiavello, D.\ Bruß, and P.\ Mataloni, Hyperentangled Mixed Phased Dicke States: Optical Design and Detection, Phys. Rev. Lett. 105, 250501 (2010).
	\bibitem{marty2} O.\ Marty, M.\ Cramer, and M.B.\ Plenio, Practical Entanglement Estimation for Spin-System Quantum Simulators Phys.\ Rev.\ Lett.\ 116, 105301 (2016).
	\bibitem{cramer2} M.\ Cramer, et al., Spatial entanglement of bosons in optical lattices, Nat.\ Comm. 4, 2161 (2013).
	\bibitem{daley} D.A.\ Abanin and E.\ Demler, Measuring Entanglement Entropy of a Generic Many-Body System with a Quantum Switch Phys.\ Rev.\ Lett.\ 109, 020504 (2012); A.J.\ Daley, et al., Measuring Entanglement Growth in Quench Dynamics of Bosons in an Optical Lattice, Phys.\ Rev.\ Lett.\ 109, 020505 (2012).
	\bibitem{islam} R.\ Islam, Measuring entanglement entropy in a quantum many-body system, Nature 528, 77 (2015).
	\bibitem{novo} L.\ Novo, T.\ Moroder, and O.\ G\"uhne, Genuine multiparticle entanglement of permutationally invariant states, Phys.\ Rev.\ A 88, 012305 (2013).
	\bibitem{redfield} J.H.\ Redfield, The Theory of Group-Reduced Distributions, Amer.\ J.\ Math.\ 49, 3 (1927).
	\bibitem{polya} G.\ P\'olya, Kombinatorische Anzahlbestimmungen für Gruppen, Graphen und chemische Verbindungen Acta Math.\ 68, 145 (1937).

	\bibitem{Kim} K.\ Kim, et al., Entanglement and Tunable Spin-Spin Couplings between Trapped Ions Using Multiple Transverse Modes, Phys.\ Rev.\ Lett.\ 103, 120502 (2009).
	\bibitem{CVX} M.\ Grant, and S.\ Boyd, CVX: Matlab software for disciplined convex programming, version 2.0 beta. http://cvxr.com/cvx, September 2013.
	
	\bibitem{yalmip} J.\ L\"ofberg, YALMIP : A toolbox for modeling and
	optimization in MATLAB, in Proc. IEEE International Symposium on Computer
	Aided Control Systems Design, Taipei, Taiwan, 2004.
	

	
\end{thebibliography}
\end{document}